\documentclass{emulateapj}
\usepackage{natbib}
\usepackage{bm}
\usepackage{mathrsfs,amsmath}
\usepackage{graphicx}
\usepackage{epstopdf}
\usepackage[english]{babel}
\usepackage[colorlinks,linkcolor={blue},citecolor={blue},urlcolor={blue}]{hyperref}
\newcommand{\mpchi}{\,h^{-1}{\rm {Mpc}}}
\newcommand{\kpchi}{\,h^{-1}{\rm {kpc}}}
\newcommand{\kms}{\,{\rm {km\, s^{-1}}}}
\newcommand{\msun}{M_{\sun}}
\newcommand{\hi}{{\rm H}{\sc i}}
\newcommand{\mhi}{M_{\mbox{H{\sc i}}}}
\newcommand{\wrp}{w_{\rm p}}
\newcommand{\rp}{r_{\rm p}}
\newcommand{\wprp}{w_{\rm p}(r_{\rm p})}

\shorttitle{\hi-Halo Mass Relation} \shortauthors{H. Guo et al.}
\begin{document}
\title{Constraining the \hi-Halo Mass Relation From Galaxy Clustering}

\author{Hong Guo\altaffilmark{1}, Cheng Li\altaffilmark{2,1}, Zheng Zheng\altaffilmark{3}, H.J.~Mo\altaffilmark{2,4}, Y.P.~Jing\altaffilmark{5,6}, Ying Zu\altaffilmark{7}, S.H.~Lim\altaffilmark{4}, Haojie Xu\altaffilmark{3}}

\altaffiltext{1}{Key Laboratory for Research in Galaxies and Cosmology, Shanghai Astronomical Observatory, Shanghai 200030, China; guohong@shao.ac.cn}
\altaffiltext{2}{Tsinghua Center for Astrophysics \& Physics Department, Tsinghua University, Beijing 100084, China; cli2015@tsinghua.edu.cn}
\altaffiltext{3}{Department of Physics and Astronomy, University of Utah, UT 84112, USA}
\altaffiltext{4}{Department of Astronomy, University of Massachusetts, Amherst MA 01003-9305, USA}
\altaffiltext{5}{Center for Astronomy and Astrophysics, Department of Physics and Astronomy, Shanghai Jiao Tong University, Shanghai 200240, China}
\altaffiltext{6}{IFSA Collaborative Innovation Center, Shanghai Jiao Tong University, Shanghai 200240, China}
\altaffiltext{7}{Center for Cosmology and AstroParticle Physics (CCAPP), Ohio State University, Columbus, OH 43210, USA}

\begin{abstract}
We study the dependence of galaxy clustering on \hi\ mass using $\sim$16,000 galaxies with redshift in the range of $0.0025<z<0.05$ and \hi\ mass of $\mhi>10^8\msun$, drawn from the 70\% complete sample of the Arecibo Legacy Fast ALFA survey. We construct subsamples of galaxies with $\mhi$ above different thresholds, and make volume-limited clustering measurements in terms of three statistics: the projected two-point correlation function, the projected cross-correlation function with respect to a reference sample, and the redshift-space monopole moment. In contrast to previous studies, which found no/weak \hi-mass dependence, we find both the clustering amplitudes on scales above a few Mpc and the bias factors to increase significantly with increasing \hi\ mass for $\mhi>10^9\msun$. For \hi\ mass thresholds below $\sim 10^9\msun$, the inferred galaxy bias factors are systematically lower than the minimum halo bias from mass-selected halo samples. We extend the simple halo model, in which the galaxy content is only determined by halo mass, by including the halo formation time as an additional parameter. A model that puts \hi-rich galaxies into halos that formed late can reproduce the clustering measurements reasonably well. We present the implications of our best-fitting model on the correlation of \hi\ mass with halo mass and formation time, as well as the halo occupation distributions and \hi\ mass functions for central and satellite galaxies. These results are compared with the predictions from semi-analytic galaxy formation models and hydrodynamic galaxy formation simulations.
\end{abstract}


\section{Introduction}

In the current paradigm of galaxy formation, baryon gas is expected to fall into the gravitational potential wells of dark matter halos, where it can cool, condense, and form stars. A key goal in modern galaxy formation is thus to understand the physical link between galaxies and their host dark matter halos. Indeed, this has been one of the key goals of the large optical imaging and spectroscopic surveys accomplished in the past one and a half decades such as the Sloan Digital Sky Survey \citep[SDSS;][]{York00}. Based on these surveys the abundance and clustering have been measured to high precision for low-redshift galaxies selected by stellar mass and multi-band luminosities  \citep[e.g.][]{Zehavi05,Li06,Li09,Yang12,Reddick13,Skibba14}. These measurements, together with various halo-based statistical models of galaxy distributions \citep[see e.g.,][]{Jing98, Berlind02, Yang03, Zheng05, Zehavi11, Guo14, Guo15c, Guo16}, have led to dramatic improvements in our understanding of the relationship between galaxy properties (e.g., stellar mass, luminosity, and color) and dark matter halos.

Compared to the stellar contents of galaxies, our understanding of their cold gas contents lags considerably behind. The gas distribution around galaxies can be used to probe the accretion, star formation and feedback processes, providing additional tests to galaxy formation models. For high-redshift galaxies, the gas properties are typically studied through analyzing the hydrogen and metal absorption lines in the spectra of background galaxies and quasars \citep[e.g.,][]{Steidel10}. At low redshifts, the gas distribution of galaxies can be mapped out through the 21cm \hi\ hyperfine emission line. Large surveys of the cold gas content of galaxies through \hi\ have become available only in the past decade. The \hi\ Parkes All-Sky Survey \citep[HIPASS;][]{Meyer04} detected $\sim5,000$ extragalactic \hi\ sources out to $z\sim0.04$ covering the whole southern sky, while the Arecibo Fast Legacy ALFA Survey \citep[ALFALFA;][]{Giovanelli05} detected more than $30,000$ extragalactic \hi\ sources out to $z\sim0.06$ in the northern sky. These surveys have enabled us to estimate the \hi\ mass function of the \hi-rich galaxies in the local universe \citep{Zwaan05,Martin10}, providing important constraints on the mass density of the atomic neutral hydrogen. However, these early surveys are limited by the effective frequency ranges and can only map the \hi\ distribution in the local universe. In the future, the Square Kilometre Array (SKA)\footnote{http://skatelescope.org/} project will provide unprecedented opportunities to probe the \hi\ distribution into much deeper universe. Indeed, the on-going Australian SKA Pathfinder (ASKAP)\footnote{http://www.atnf.csiro.au/projects/askap/} survey and MeerKAT\footnote{http://www.ska.ac.za/science-engineering/meerkat/} science projects will provide a large amount of \hi\ data in the forthcoming years. For example, the Wide-field ASKAP L-Band Legacy All-Sky Blind Survey \citep[WALLABY;][]{Koribalski12} will detect up to $500,000$ \hi-selected galaxies to a redshift of $z\sim0.26$, the Westerbork Northern Sky \hi\ Survey \citep[WNSHS;][]{Duffy12b} is targeting at the northern sky complimentary to WALLABY,  the Deep Investigations of Neutral Gas Origins \citep[DINGO;][]{Meyer09} survey will study the evolution of the gas-rich galaxies to $z\sim0.4$, the COSMOS HI Large Extragalactic Survey \citep[CHILES;][]{Fernandez13,Fernandez16} will probe the \hi\ deep field centered at the COSMOS filed to $z\sim0.45$, and the Looking at the Distant Universe with the MeerKAT Array survey \citep[LADUMA;][]{Holwerda12} is designed to detect \hi\ to $z>1$. 

Galaxy clustering provides a way to help establish the connection between the \hi\ content of galaxies and dark matter halos. A commonly used statistic for quantifying the galaxy clustering is the two-point correlation function (2PCF), which measures the excess probability of finding pairs of galaxies at certain separations with respect to a random distribution. The 2PCF has been measured for the HIPASS and ALFALFA samples \citep{Basilakos07,Meyer07,Martin12,Papastergis13}. The \hi-selected galaxies are found to have low clustering amplitudes, with a linear galaxy bias, $b_{\rm g}$, varying from 0.7 to 0.9 for different galaxy samples. However, the previous measurements of the clustering dependence on the \hi\ mass is still controversial. Based on the same HIPASS galaxy sample, \cite{Basilakos07} found a much stronger clustering amplitude for galaxies with a larger \hi\ mass, while \cite{Meyer07} found only very weak dependence.  By using the 40\% complete catalog of the ALFALFA survey, \cite{Papastergis13} claimed to find no strong dependence of the clustering amplitudes of \hi-selected galaxies on the \hi\ mass over the range of $\mhi=10^{8.5}\sim10^{10.5}\msun$. However, we note that there is a trend of \hi\ mass dependence in their measurements of \hi\ mass bin samples (see their Figure~10), but the authors attributed it to the finite volume effect in the small ALFALFA sample.

It has been found that the \hi\ mass generally increases with galaxy stellar mass, and that the stellar mass in this \hi\ mass range would vary from $10^7$ to $10^{12}\msun$ \citep{Bothwell09,Zhang09,Zhang12,Evoli11,Saintonge11,McGaugh12,Popping15}. Given the strong dependence of galaxy clustering on stellar mass as well established in recent studies \citep[see e.g.,][]{Li06,Li09,Yang12,Reddick13,Skibba14}, the no/weak dependence of clustering on \hi\ mass is puzzling, even if a reasonable scatter between the \hi\ and stellar masses is taken into account to weaken the dependence of clustering on \hi\ mass.

Accurate clustering measurements of the \hi-selected galaxies are important in constraining the \hi-halo mass relation, as the galaxy bias is directly related to their host halo masses. The \hi-halo mass relation has been studied in hydrodynamical simulations \citep[see e.g.,][]{Duffy12,Dave13,Cunnama14,Vogelsberger14b,Rafieferantsoa15,Crain17} and semi-analytical models \citep[see e.g.,][]{Blitz06,Obreschkow09,Popping09,Popping14b,Fu10,Fu12,Fu13,Lagos11b,Kim17,Xie17,Zoldan17}. The predicted \hi-halo mass relations from the various models are highly dependent on the detailed algorithms and have significant differences. In addition, semi-empirical models have been developed to match the observed \hi\ mass functions \citep[see e.g.,][]{Popping15,Padmanabhan17,Padmanabhan17a,Padmanabhan17b}. While the \hi-halo mass relation may be reliably inferred from the spatial clustering of the observed \hi-selected galaxies using halo-based statistical models, it has not been well studied so far. Modeling of the \hi-halo mass relation would potentially be able to provide new and interesting constraints on galaxy formation models.

In this paper, we use the 70\% complete catalog of the \hi-selected galaxies in ALFALFA to examine the dependence of clustering on \hi\ mass, as well as to investigate the statistical relation between \hi\ mass and halo mass. When estimating the 2PCFs we carefully take into account the effect of the ALFALFA sample selection, which depends on both the flux limit and the \hi\ emission line width, as well as the sample variance effect caused by the limited survey volume. This gives rise to the finding of a significant dependence of clustering on the \hi\ mass. We extend the simple sub-halo abundance matching (SHAM; e.g., \citealt{Conroy06}) model to interpret the observed clustering of the different \hi\ mass samples. Different from the SHAM model adopted in the literature which links the stellar mass or luminosity of galaxies with the maximum circular velocity or mass of the host halos, our model additionally takes into account the dependence of clustering on halo formation time. We will show that the inclusion of the halo formation time in the modeling is required by the data, enabling our model to successfully fit the observed \hi\ mass dependence of the clustering and thus constrain the \hi-halo mass relation.

The structure of the paper is constructed as follows. In \S\ref{sec:data}, we describe the galaxy samples and the simulation used in the modeling. The clustering measurements are displayed in \S\ref{sec:measurements}. We introduce our modeling method in \S\ref{sec:model}. The model implications are presented in \S\ref{sec:hihm}. We summarize and discuss our results in \S\ref{sec:conclusion} and \S\ref{sec:discussion}, respectively. Throughout the paper, we assume a spatially flat $\Lambda$ cold dark matter cosmology, with $\Omega_{\rm m}=0.307$, $h=0.678$, $\Omega_{\rm b}=0.048$ and $\sigma_8=0.823$, consistent with the constraints from Planck \citep{PlanckCollaboration14} and with the parameters used in the simulation adopted for our modeling (see \S\ref{sec:model}). 

\section{Data} \label{sec:data}
\begin{table}
	\caption{Samples of different \hi-mass thresholds} \label{tab:sample}
	\centering
	\begin{tabular}{lrr}
		\hline
		Sample  & $N_{\rm gal}$ & $n_{\rm g}(h^{3}{\rm {Mpc}}^{-3})$  \\
		\hline		
		$\log(\mhi/M_{\odot})>8.0$  & 16059 & $ 136.49\times10^{-3}$ \\
		$\log(\mhi/M_{\odot})>8.2$  & 15900 & $ 113.52\times10^{-3}$ \\
		$\log(\mhi/M_{\odot})>8.4$  & 15684 & $ 89.86\times10^{-3}$ \\
		$\log(\mhi/M_{\odot})>8.6$  & 15395 & $ 68.97\times10^{-3}$ \\
		$\log(\mhi/M_{\odot})>8.8$  & 14989 & $ 51.95\times10^{-3}$ \\
		$\log(\mhi/M_{\odot})>9.0$  & 14394 & $ 39.05\times10^{-3}$ \\
		$\log(\mhi/M_{\odot})>9.2$  & 13363 & $ 26.98\times10^{-3}$ \\
		$\log(\mhi/M_{\odot})>9.4$  & 11752 & $ 17.50\times10^{-3}$ \\
		$\log(\mhi/M_{\odot})>9.6$  & 9555 & $ 10.51\times10^{-3}$ \\
		$\log(\mhi/M_{\odot})>9.8$  & 6775 & $ 5.84\times10^{-3}$ \\
		$\log(\mhi/M_{\odot})>10.0$  & 3841 & $ 2.68\times10^{-3}$ \\
		$\log(\mhi/M_{\odot})>10.2$  & 1444 & $ 0.92\times10^{-3}$ \\
		$\log(\mhi/M_{\odot})>10.4$  & 353 & $ 0.22\times10^{-3}$ \\			
		\hline
	\end{tabular}
	
	\medskip
	All the \hi-mass samples cover the same redshift range of $0.0025<z<0.05$, and the effective comoving volume is $1.52\times10^6(h^{-1}\,\rm{Mpc})^3$. The total number of galaxies $N_{\rm gal}$ and the mean number density $n_g$ are also displayed (see Equation~(\ref{eq:ng})).
\end{table}
\subsection{The ALFALFA survey and the HI galaxy sample}
In this paper, we use the most up-to-date public data release, $\alpha.70$\footnote{http://egg.astro.cornell.edu/alfalfa/data/index.php}, of the ALFALFA survey \citep{Giovanelli05}, which is a catalog of 70\% of the final survey area. The blind observation of the 21-cm emission line is performed with the 305-m single-dish radio telescope at the Arecibo Observatory, which is very sensitive in the L-band of 1.4~GHz with an angular resolution of $3.5^\prime$. The optical counterparts of these \hi\ sources are identified interactively by cross-matching external imaging databases \citep[see details in][]{Haynes11}. For our analysis, we only select \hi\ detections with secure extragalactic sources, i.e., ``Code 1'' in the ALFALFA catalog. An \hi\ mass is calculated for each \hi\ source in the catalog from the integrated flux density of the \hi\ emission line and the galaxy luminosity distance \citep[see Eq. 1 of][]{Giovanelli05}.

\begin{figure*}
\centering
\includegraphics[width=0.8\textwidth]{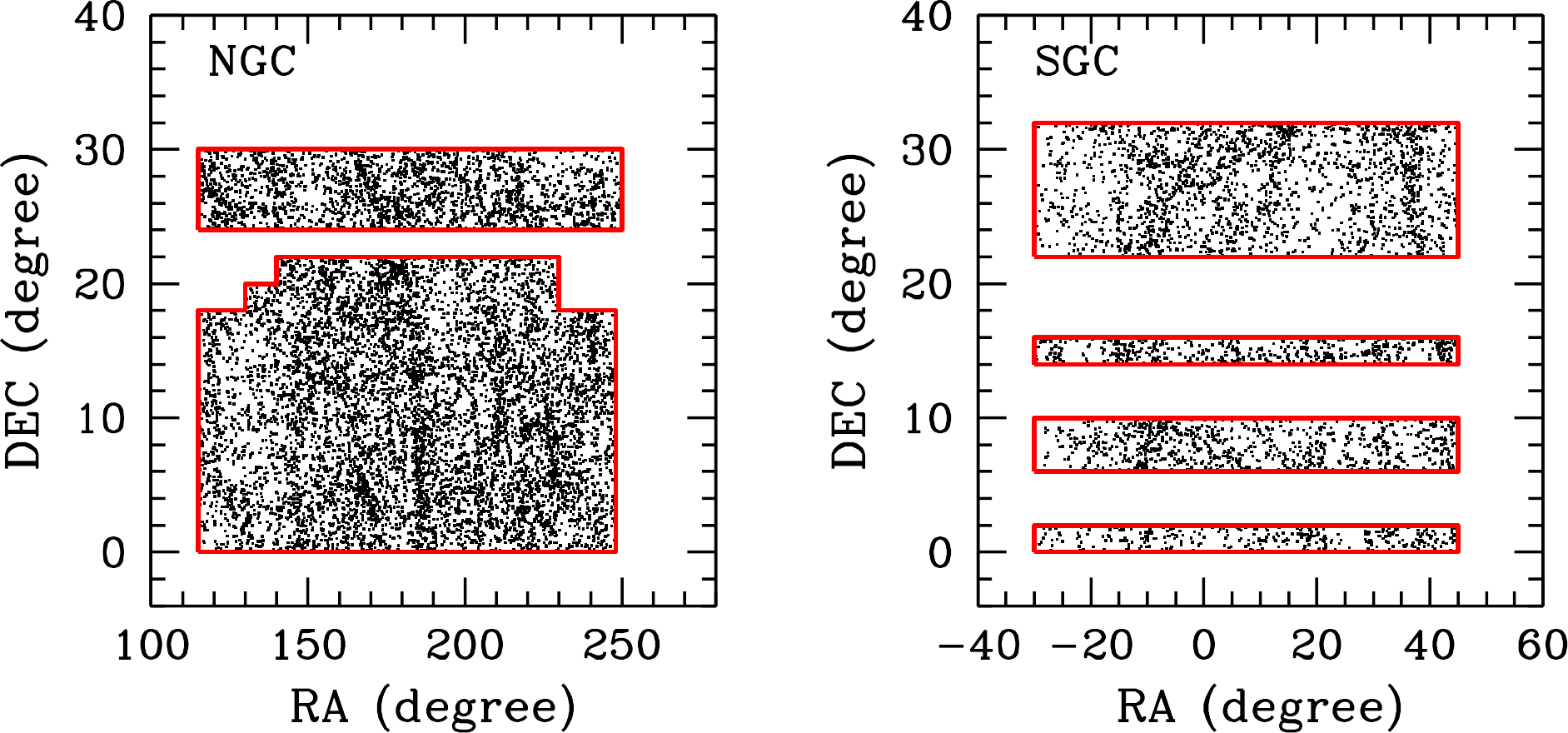}
\caption{Angular distribution of galaxies in the ALFALFA $\alpha.$70 sample, separated into the northern (left) and southern (right) galactic caps. For simplicity, we only select galaxies in the ALFALFA survey limited to the simple RA and DEC cuts shown as the red boundaries.
} \label{fig:radec}
\end{figure*}

Although ALFALFA is a blind survey, the resulting sample of the \hi-detected sources is not purely flux-limited. The detection rate depends on both the \hi\ line flux, $S_{\rm int}$, and the line profile width, $W_{50}$, as the detector is more sensitive to narrower line profiles than broader ones for a given $S_{\rm int}$. We apply the completeness cut suggested by \cite{Haynes11} (see their Eqs. 4 and 5) to select a sample that is more than 50\% complete. Figure~\ref{fig:radec} displays the angular distribution of the ALFALFA sample, where the right ascension (RA) and declination (DEC) are the angular positions of the optical counterparts of the \hi\ sources. For simplicity, we only select galaxies in the ALFALFA survey limited to the RA and DEC cuts shown as the red boundaries in the figure.

The redshift of each source in the ALFALFA catalog corresponds to the heliocentric velocity $v_{21}\equiv cz_{21}$ of the \hi\ 21-cm emission line. We convert $z_{21}$ to the redshift in the cosmic microwave background frame \citep{Fixsen96}, $z_{\rm CMB}$, for the clustering measurements. Because of the radio frequency interference (RFI) from the Federal Aviation Administration radar at high heliocentric velocities, we further limit the redshift range to $0.0025<z_{\rm CMB}<0.05$. Other sources of RFI still contaminate the signals in regions of frequency space corresponding to spherical shells in the survey volume. The average weights of the lost volume due to RFI, $w_{\rm RFI}$, has been estimated for galaxies with different heliocentric velocities in ALFALFA \citep[see Fig.~6 of][]{Martin10}. We will correct for this effect using the random galaxy catalogs, as will be discussed in \S~\ref{sec:method}.

These restrictions produce a sample of 16,313 \hi-detected galaxies in the redshift range of $0.0025<z<0.05$ and with reliable estimates of the 
\hi\ mass ranging from $10^{7.1}\msun$ to $10^{10.9}\msun$. The sample covers a total area of 4,693 ${\rm deg}^2$ in the sky and an effective comoving volume of $1.52\times10^6\,{\rm Mpc}^3$.

\subsection{\hi-mass samples}
\begin{figure*}
	\centering
	\includegraphics[width=0.8\textwidth]{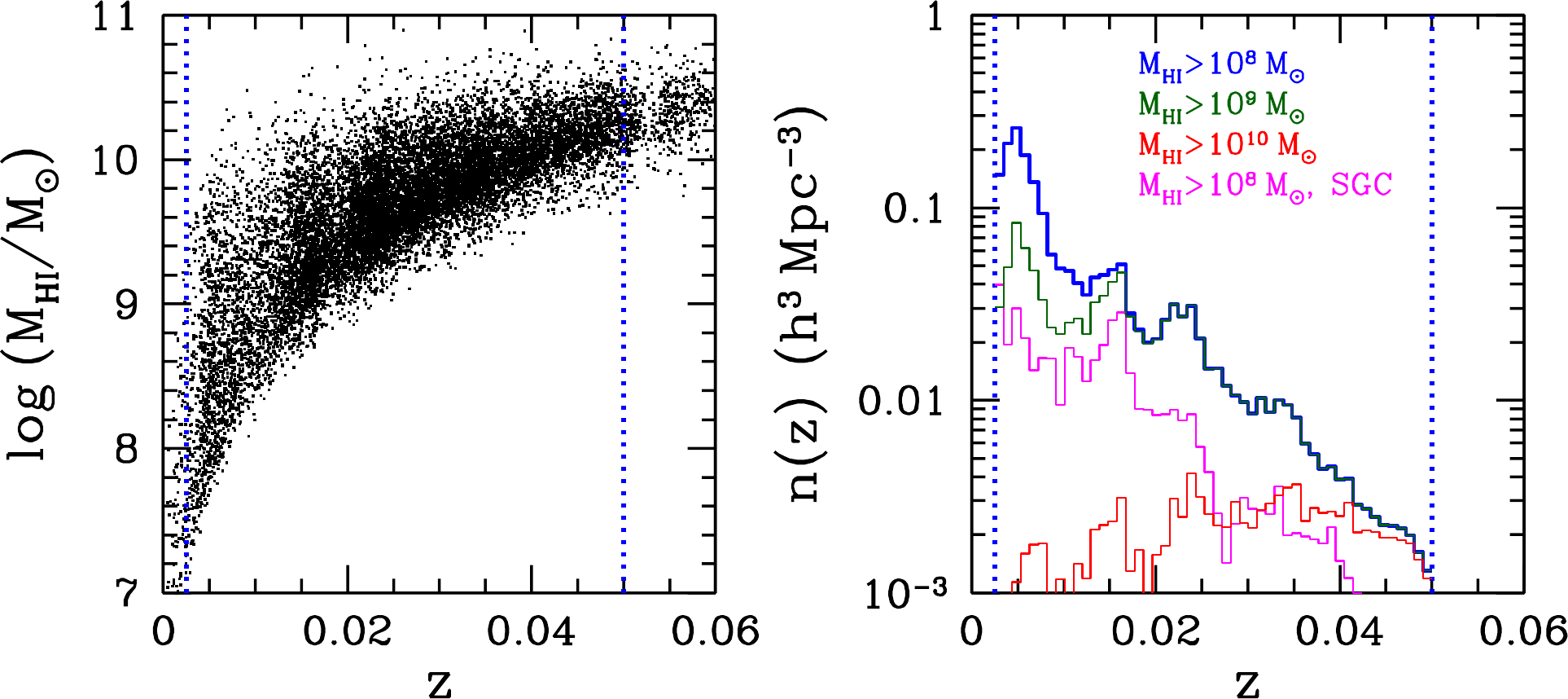}
	\caption{Left: distribution of the \hi-selected galaxies as a function of redshift $z_{\rm CMB}$ and \hi\ mass $\mhi$. The two blue vertical lines show the redshift limits of our sample. Right: number density distribution for three typical $\mhi$ samples, with the blue, green and red histograms for $\mhi>10^8\msun$, $10^9\msun$ and $10^{10}\msun$, respectively. The number density distribution for the $\mhi>10^8\msun$ sample in the southern galactic cap is shown as the magenta histogram.
	} \label{fig:selection}
\end{figure*}
In order to study the dependence of clustering on \hi\ mass, we have constructed a set of 13 \hi\ mass samples selected by using different minimum masses, with a fixed mass threshold interval of $\Delta\log(\mhi/M_{\odot})=0.2$. Table~\ref{tab:sample} lists the detailed sample information, including the total number of galaxies and the average number density (calculated using Equation~(\ref{eq:ng})) of each sample. 

The left panel of Figure~\ref{fig:selection} shows the distribution of the \hi-selected galaxies in redshift ($z$) - \hi\ mass ($\mhi$) plane. The two blue vertical lines indicate the redshift limits of our total sample. The apparent decrease in the number of galaxies around $z\sim0.052$ is caused by the aforementioned RFI. The right panel presents the number density distribution $n(z)$ for three typical $\mhi$ threshold samples, with the blue, green, and red histograms for $\mhi>10^8\msun$, $10^9\msun$ and $10^{10}\msun$, respectively. The bumps and dips indicate the influence of local super-clusters. For instance, the bump at $z\sim 0.005$ reflects the significant spatial overdensity of galaxies with $\mhi>10^8\msun$ in the Virgo Cluster. However, galaxies in the Virgo Cluster are generally \hi-deficient when compared to field galaxies of the same size and morphological type \cite[see e.g.,][]{Solanes01,Gavazzi05,Chung09}. The dip at $z\sim 0.02$ is caused by the fact that only very few galaxies in the Coma Cluster have $\mhi>10^9\msun$ \citep{Giovanelli85,Magri88,Bravo-Alfaro00}. These large-scale structures have a significant effect on the \hi\ distribution in the local volume, especially for the low $\mhi$ samples, as we will see below. This is actually one of the main reasons why we opt for \hi-mass thresholds instead of differential mass bins, as the latter samples will be significantly affected by the limited volumes. \hi-mass threshold samples have the largest available volume of the ALFALFA survey, minimizing the sample variance effect. 

In the figure we additionally show the number density distribution of the $\mhi>10^8\msun$ sample, but only for the subset in the southern galactic cap. Both the peak at $\sim0.005$ and the dip at $\sim0.02$ become less prominent in comparison to the full sample at the same mass threshold, indicating that the 
northern galactic cap (NGC) is indeed overwhelmingly dominated by the super-clusters including Virgo and Coma. 

The drop in the number density for each of the $\mhi>10^8\msun$ and $10^9\msun$ samples toward higher redshift reflects the fact that the sample is flux-limited, not volume-limited (see the left panel). We note that even though each \hi\ mass threshold sample is not volume-limited, the clustering measurements will be made in an effectively volume-limited sense, as detailed in \S~\ref{sec:method}.

\subsection{SDSS/DR7 galaxy sample}

Our analysis also makes use of the Sloan Digital Sky Survey \citep[SDSS;][]{York00} Data Release 7 \citep[DR7;][]{Abazajian09} Main galaxy sample, to investigate the correlation between the \hi\ sources and the general population of galaxies. The ALFALFA survey covers much larger area than the SDSS/DR7 in the southern galactic cap (SGC), but the two samples overlap significantly in the northern galactic cap (NGC). We take the SDSS/DR7 galaxy data from the New York University Value-Added Galaxy Catalog \citep[NYU-VAGC;][]{Blanton05c}, where the effective area of the sample is about $7,300\deg^2$ and the galaxies are selected simply by an $r$-band Petrosian magnitude limit of $r\sim 17.77$. For each galaxy the k+e corrected luminosities in the SDSS $ugriz$ bands and a stellar mass estimated from the SDSS photometry are also provided in this catalog \citep{Blanton07b}.

\subsection{The SMDPL simulation}

We use catalogs of dark matter halos and subhalos identified from the Small MultiDark simulation of Planck cosmology (SMDPL\footnote{The SMDPL halo and subhalo catalogs are publicly available at http://dx.doi.org/10.17876/cosmosim/smdpl/}; \citealt{Klypin16}) to model the statistical relation between the \hi-selected galaxies and host dark matter halos. The SMDPL simulates the evolution of dark matter distribution in a comoving volume of $400^3\,h^{-3}$\,Mpc$^3$ with a mass resolution of $9.6\times10^7h^{-1}\msun$, assuming cosmological parameters of $\Omega_m=0.307$, $\Omega_b=0.048$, $h=0.678$, $n_s=0.96$, and $\sigma_8=0.823$. The dark matter halos and subhalos in SMDPL are identified with the \texttt{ROCKSTAR} phase-space halo finder \citep{Behroozi13c}, which is shown to be efficient and accurate to find the bound spherical structures \citep{Onions12,Knebe13}. For the current work we use the simulation output at $z=0$ which is close to the average redshift of the ALFALFA survey. 

\section{Measuring the clustering and biasing of \hi-mass selected samples}
\label{sec:measurements}

\subsection{Methodology of clustering measurements}
\label{sec:method}

\begin{figure*}
	\centering
	\includegraphics[width=\textwidth]{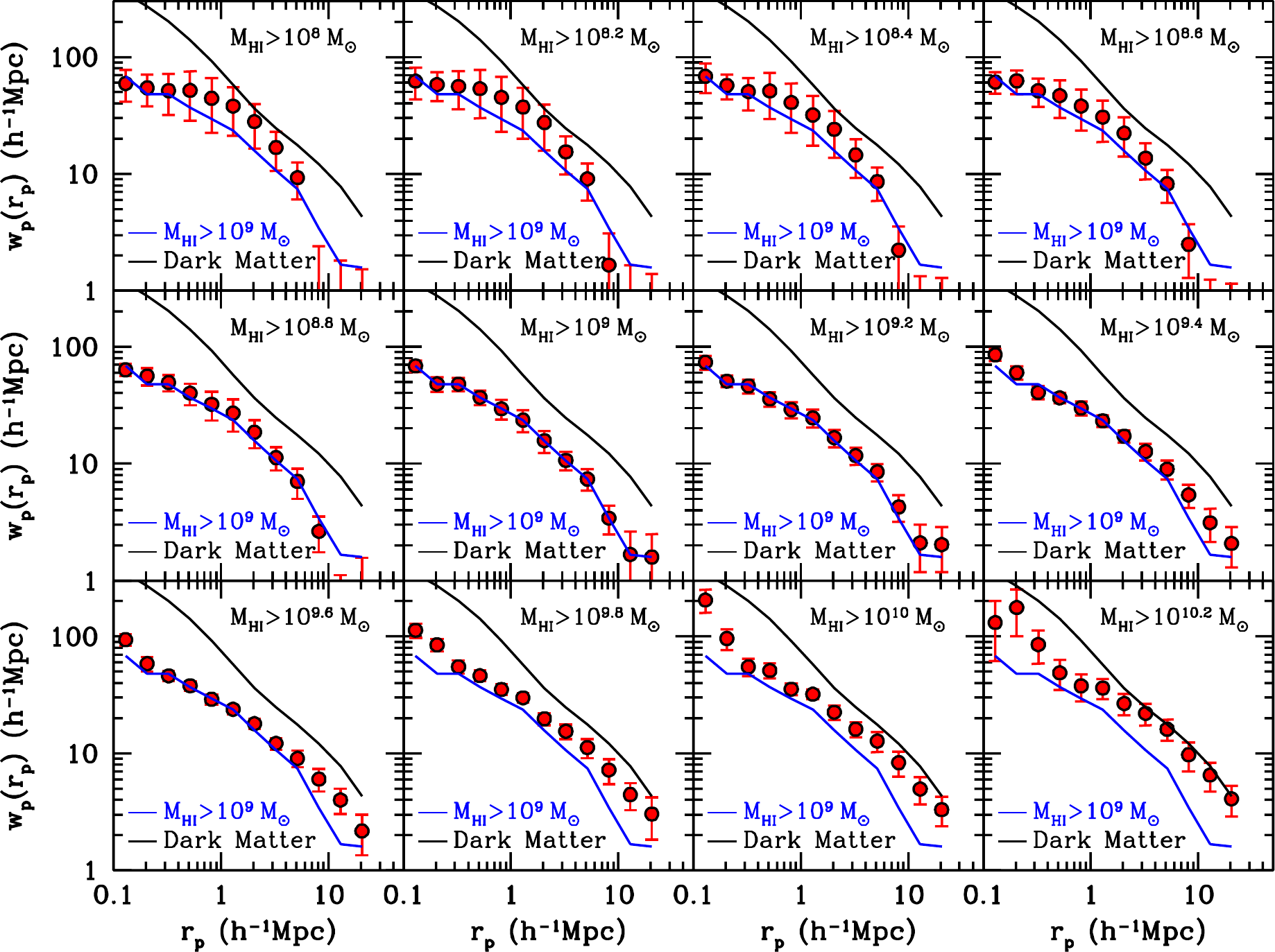}
	\caption{Projected two-point correlation function for 12 of the 13 \hi-mass threshold samples. In each panel, the symbols with error bars present $w_{\rm p}$ as a function of $r_{\rm p}$ and its error for a given \hi-mass threshold sample. For comparison, the measurement of the sample with $\mhi>10^9\msun$ is repeated in every panel as a blue solid line. We also show the predicted dark matter distribution as the black line.
	}
	\label{fig:wp_MHI}
\end{figure*}

As described in the previous section, when selecting the ALFALFA galaxy sample we have applied the completeness limit, which depends on the \hi\ line profile. Therefore, each galaxy in the catalog has a different flux limit $S_{\rm int, lim}$, depending on the width of its \hi\ line profile. Different from galaxy clustering analyses that use volume-limited samples in SDSS DR7 \citep[see e.g.,][]{Zehavi11,Guo15b}, constructing volume-limited samples for low-$\mhi$ galaxies in ALFALFA is impractical due to the limited redshift range and the $W_{50}$-dependent flux limit. Instead, we measure the clustering of each sample using galaxies in the whole redshift range and correct for the effect of $S_{\rm int, lim}$ following the method laid out in \cite{Xu16}. Such a method makes use of the maximum volume $V_{\rm max}$ accessible to each galaxy, leading to an effectively volume-limited clustering measurement from a flux-limited sample \citep{Xu16}. 

Given the small volume of the ALFALFA survey, the local large-scale structures can potentially affect the clustering measurements even if we use \hi\ mass threshold samples over a relatively large redshift range. We apply the maximum-likelihood method \citep{Zwaan05} to calculate the effective volume available to each galaxy $V_{\rm eff}$ as the value of $V_{\rm max}$, which takes into account the survey sensitivity limit and the density fluctuations of galaxies caused by the large-scale structure. The details of the method can be found in appendix B of \cite{Martin10} \citep[see also][]{Papastergis11}. This has proved very robust against density fluctuations along the line of sight \citep{Jones16}.

For each of our \hi-mass threshold samples, we use a generalized Landy-Szalay estimator \citep{Landy93} to measure the  redshift-space 2PCFs $\xi(r_{\rm p},r_{\rm\pi})$ \citep{Li09,Xu16}, 
\begin{equation}
\xi(r_{\rm p},r_{\rm\pi})=\frac{\rm{DD^*-2DR^*+RR^*}}{\rm{RR^*}}, \label{eq:xi}
\end{equation}
where $r_{\rm\pi}$ and $r_{\rm p}$ are the separations of galaxy pairs along and perpendicular to the line-of-sight (LOS). The data-data ($\rm{DD^*}$), data-random ($\rm{DR^*}$), and random-random ($\rm{RR^*}$) pair counts are calculated as follows,
\begin{eqnarray}
\rm{DD}^*(r_{\rm p},r_{\rm\pi})&=&\sum_{(i,j)\in V_{ij}}\frac{1}{V_{ij}} \label{eq:dd}, \\
\rm{DR}^*(r_{\rm p},r_{\rm\pi})&=&\sum_{(i,j)\in V_{ij}}\frac{1}{V_{ij}} \label{eq:dr},\\
\rm{RR}^*(r_{\rm p},r_{\rm\pi})&=&\sum_{(i,j)\in V_{ij}}\frac{1}{V_{ij}}, \label{eq:rr}
\end{eqnarray}
where $V_{ij}={\rm min}(V_{{\rm eff},i},V_{{\rm eff},j})$, with $V_{{\rm eff},i}$ and $V_{{\rm eff},j}$ being the effective volumes accessible to the $i$-th and $j$-th objects, respectively. Note that in Eqs.~(\ref{eq:dd})--(\ref{eq:rr}) we only include pairs in which both of the galaxies are within the common volume $V_{ij}$. As stated in the Appendix of \cite{Xu16}, at small $r_p$ separations, such a requirement would possibly exclude a small fraction of correlated pairs stretched along the line of sight by the effect of peculiar velocities. By comparing to the case of including all galaxy pairs, we quantify this effect to be much smaller than 2\% for $\rp<1\mpchi$, which is negligible compared to the measurement errors at these scales. For simplicity, we assume, in the above expressions, that the number of galaxies in the random catalog is the same as that in the data catalog. In practice, the random catalog we construct (see below) for each sample is 50 times as large as the data catalog, and all the pair counts are correctly normalized. The generalized Landy-Szalay estimator weighs each galaxy pair according to the common maximum volume that the pair of galaxies can both be observed, i.e., the contribution of each pair to the 2PCF measurement is accounted for in a volume-limited sense. In the end, we achieve an effectively volume-limited 2PCF measurement by making full use of a flux-limited sample \citep{Xu16}. Compared to a traditional volume-limited sample, the sample variance is reduced in our approach.

The average sample number density $n_{\rm g}$ is estimated as
\begin{equation}
n_{\rm g}=\sum_i \frac{1}{V_{{\rm eff},i}} \label{eq:ng}.
\end{equation}
This number density is substantially larger than that computed using the number of galaxies listed in Table~\ref{tab:sample} and the sample comoving volume ($1.55\times10^6\,h^{-3}\,\rm{Mpc}^3$), as the sample is essentially flux-limited, which leads to larger incompleteness at larger redshifts.

The random samples are constructed in the following way. First, we generate a large set of random points uniformly distributed in the RA and $\sin(\rm{DEC})$ plane within the survey area. The discontinuity in the angular distribution of the $\alpha$.70 sample (especially the SGC) is taken into account in the random sample with the same angular geometry. We then assign each random point a set of properties (redshift and \hi\ mass, as well as the corresponding $V_{\rm eff}$), which are the same as those of a real galaxy randomly drawn from the full sample of \hi\ galaxies. This method has been verified to be reliable and accurate for wide-angle surveys such as SDSS and ALFALFA \citep[see e.g.,][]{Li06, Ross12}. As the redshifts come directly from the full galaxy sample, the RFI effects seen in the data are automatically built into the random catalog. Next, for each \hi-mass threshold galaxy sample, we construct the corresponding random sample by selecting from the full random catalog the points with \hi\ masses above the threshold. 

To reduce the effect of redshift-space distortion (RSD) caused by galaxy peculiar velocities, we focus on the measurements of the projected 2PCF $w_{\rm p}(r_{\rm p})$ \citep{Davis83}, defined as
\begin{equation}
w_{\rm p}(r_{\rm p})=\int_{-r_{\pi,{\rm max}}}^{r_{\pi,{\rm max}}} \xi(r_{\rm p},r_{\rm\pi})dr_{\rm\pi}. \label{eq:wp}
\end{equation}
In practice, the integration runs to a limited line-of-sight separation $r_{\pi,{\rm max}}$ instead of infinity in order to reduce the noise at very large separations. To select a reasonable $r_{\pi,{\rm max}}$, we have investigated the variation of $\xi(r_{\rm p},r_{\rm\pi})$ with $r_{\rm\pi}$  for different \hi\ mass threshold samples and found that the value of $\xi(r_{\rm p},r_{\rm\pi})$ is close to zero and becomes noisy for most of the samples when $r_{\rm\pi}$ is larger than $20\mpchi$. Therefore, to maximize the signal to noise ratio (S/N) of the $w_{\rm p}$ measurement we adopt $r_{\pi,{\rm max}}=20\mpchi$, and the same value is used in our models. We adopt logarithmic $r_{\rm p}$ bins 
of a constant width $\Delta\log r_{\rm p}=0.2$ covering the range from $0.13$ to $20.48\mpchi$,  
and linear $r_{\rm\pi}$ bins of width $\Delta r_{\rm\pi}=2\mpchi$ from 0 to 20$\mpchi$.
The error covariance matrices for $w_{\rm p}(r_{\rm p})$ are estimated using the jackknife re-sampling technique 
with 121 subsamples \citep{Zehavi11,Guo15b}.

\subsection{Dependence of clustering on \hi\ mass}

Figure~\ref{fig:wp_MHI} shows the $w_{\rm p}(r_{\rm p})$ measurements (symbols with error bars) for 12 of the 13 \hi-mass threshold samples. For comparison, the measurement of the sample with $\mhi>10^9\msun$ is repeated in every panel as a blue solid line. The black line is the predicted $w_{\rm p}(r_{\rm p})$ for the underlying dark matter distribution at $z=0$ assuming the same cosmology. Compared to the dark matter distribution, all the galaxy samples except the one with $\mhi>10^{10.2}\msun$ are significantly anti-biased, showing lower clustering amplitudes at all scales probed. This result is in agreement with previous studies of the clustering of \hi-rich galaxies \citep{Basilakos07,Meyer07,Li12,Martin12,Papastergis13}. 

In contrast to previous results of no dependence of clustering on \hi\ mass \citep[e.g.,][]{Meyer07,Papastergis13}, we find the clustering amplitude of the \hi-selected galaxies to depend on the \hi\ mass, in different ways at different scales, when the mass threshold exceeds $10^9\msun$. On scales larger than a few Mpc, the clustering amplitude at a given scale shows no/little change as the \hi\ mass threshold goes from the lowest value of $10^8\msun$ up to $10^9\msun$, before increasing significantly at higher masses. In the figure, the amplitude of $w_{\rm p}$ at $\sim10\mpchi$ for the most massive sample (with $\mhi>10^{10.2}\msun$) is a factor of $\sim4$ higher than that of the sample with $\mhi>10^{9}\msun$ and is comparable to that of the dark matter. It is interesting to note that the slope of the correlation function also presents a systematic trend with \hi\ mass when the threshold exceeds $10^9\msun$, with a flatter shape at higher masses. This effect makes the mass dependence more pronounced at larger scales. For instance, at $r_{\rm p}\sim10\mpchi$ the increasing amplitude is observed for all the samples with mass threshold above $10^9\msun$, but at $r_{\rm p}\sim1\mpchi$ the mass dependence becomes obvious only when $\mhi>10^{9.8}\msun$ or higher.

On scales smaller than a few Mpc, the $w_{\rm p}(r_{\rm p})$ measurement shows no significant dependence on \hi\ mass, except for intermediate-to-high mass samples with thresholds above $10^{9.8}\msun$. We note that the clustering amplitudes for samples with the \hi\ mass threshold below $10^9\msun$ appear to be enhanced on intermediate scales, from a few $\times100$ kpc to a few $\mpchi$, when compared to the samples at higher masses. These differences are significant, but they are likely produced by the strong sample variance effect due to the small volume of the samples. Although our method of weighing pairs by $1/V_{\rm max}$ reduces the sample variance effect in comparison to the measurements with the traditional volume-limited sample, the effect may still be non-negligible, as the number of galaxies with $10^8\msun<\mhi<10^9\msun$ is only 1665 (see Table~\ref{tab:sample}) and most of these galaxies are located at $z<0.015$, where the local super-clusters could dominate the clustering power (Figure~\ref{fig:selection}). We will discuss the effect of the limited volume in more detail in the next subsection.
  
\subsection{Robustness of the clustering measurements} \label{sec:test} 
Sample variance is the dominant source of uncertainty in clustering measurements when the survey volume is small. This is indeed the case for our low-mass samples, as indicated by the large error bars of the $w_{\rm p}(r_{\rm p})$ measurements for the samples with \hi\ mass thresholds below $\sim10^9\msun$ (see Figure~\ref{fig:wp_MHI}). Our measurements on large scales and at high masses may also be affected to a certain extent by the sample variance. In this subsection, we perform four analyses to test the robustness of both the clustering measurements and the error estimates. The first three analyses are based on real samples from the ALFALFA and SDSS, while the fourth analysis uses a large set of mock catalogs constructed from a cosmological simulation. With these analyses we aim to better understand the reason behind the intermediate-scale enhancement in the correlation function of low-mass samples, as well as to further verify the mass dependence of the correlation function at $\mhi>10^9\msun$.

\subsubsection{Finite Volume Effect}
\begin{figure*}
	\centering
	\includegraphics[width=0.9\textwidth]{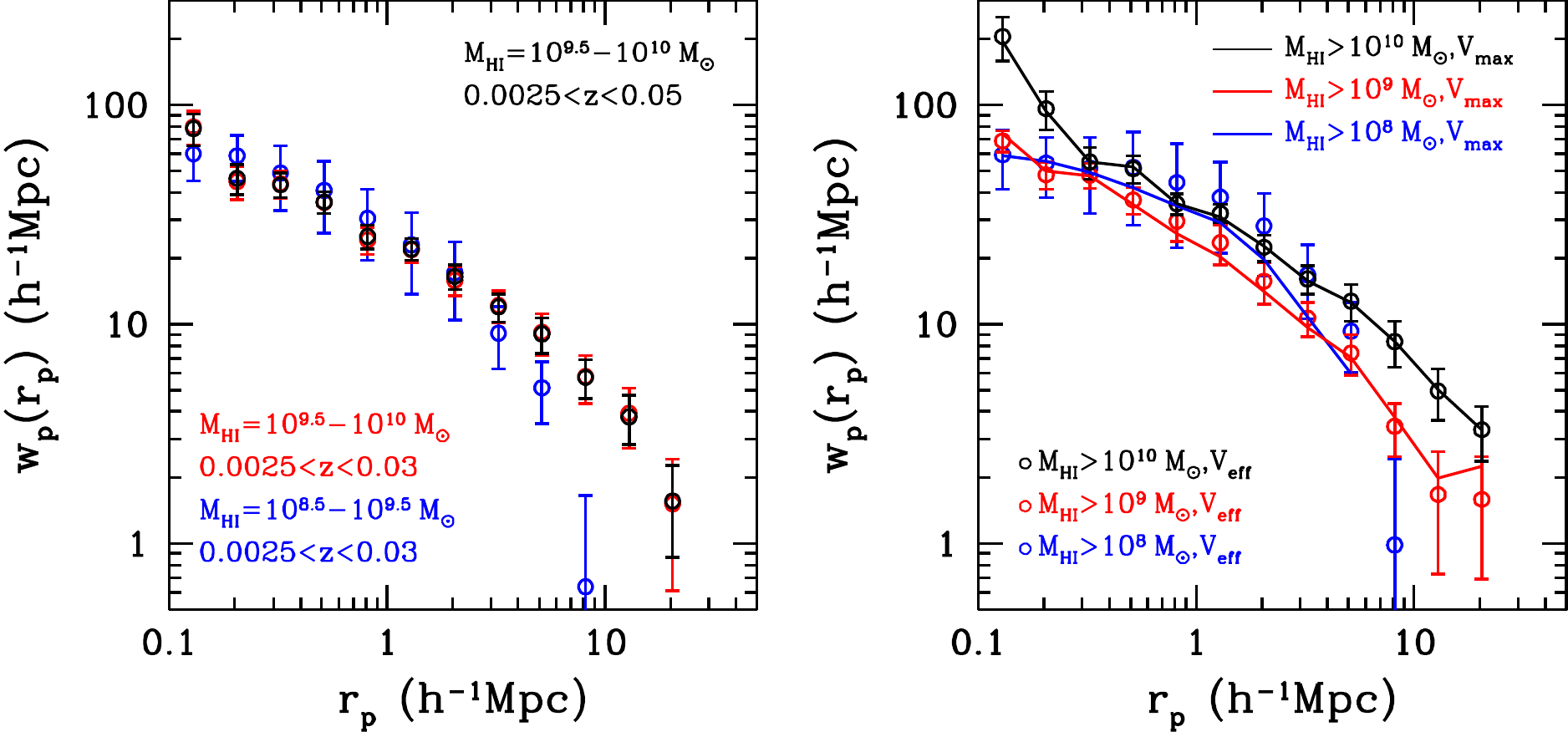}
	\caption{Left: comparisons of the $\wprp$ measurements with the \hi\ mass in the ranges of $10^{8.5}$--$10^{9.5}\msun$ (blue circles) and $10^{9.5}$--$10^{10}\msun$ (red circles) in a smaller volume of $0.0025<z<0.03$. We also show the measurements of the sample $10^{9.5}$--$10^{10}\msun$ in the redshift range of $0.0025<z<0.05$ for comparison (black circles). Right: comparisons of the $\wprp$ measurements by applying weights according to $V_{\rm max}$ and $V_{\rm eff}$, respectively (see the text).
	} \label{fig:wp_voltest}
\end{figure*}
We note that \cite{Papastergis13} found a trend of the clustering dependence on the \hi\ mass by comparing the projected 2PCFs between two samples with the \hi\ mass in the ranges of $10^{8.5}$--$10^{9.5}\msun$ and $10^{9.5}$--$10^{10}\msun$ (their Figure~10). But when limiting the higher \hi\ mass sample to the same volume as lower one, the trend disappears, which they attribute to the finite volume effect. With the $\alpha$.70 sample that is about $2.8$ times the volume as in \cite{Papastergis13}, we perform the same test with the same \hi\ mass ranges to verify our results, which is shown in the left panel of Figure~\ref{fig:wp_voltest}. We compare the measurements of the two \hi\ mass samples in a smaller volume of $0.0025<z<0.03$ (blue and red circles). The measurement for the higher \hi\ mass sample in $0.0025<z<0.05$ is also shown for comparison.

There are two important features in the figure. Firstly, in this smaller volume, we also find a clear trend of \hi\ mass dependence on large scales between the low and high mass samples, consistent with Figure~\ref{fig:wp_MHI}. Secondly, the $\wprp$ measurements for the high mass sample is not significantly affected by the finite volume effect. From the redshift--\hi\ mass distribution in Figure~\ref{fig:selection}, the high mass sample in $0.0025<z<0.03$ is almost volume-limited, if not considering the dependence on $W_{50}$. The consistency between the measurements of the high mass sample in different volumes further verifies our method of using $V_{\rm eff}$ to correct for the sample selection effect. Therefore, the detection of \hi-mass dependence of the clustering can be attributed to both the larger volume of our sample and the correction for the flux limit in the measurements.

In this paper, we use the effective volume $V_{\rm eff}$ determined from the 2D step-wise maximum likelihood method of \cite{Zwaan05}, rather than the normal maximum volume $V_{\rm max}$ from the survey completeness limit. The use of $V_{\rm eff}$ helps obtain more accurate \hi\ mass number density, which is necessary for the halo modeling. We show in the right panel of Figure~\ref{fig:wp_voltest} the measurements of $\wprp$ for three typical \hi\ mass samples using the weights of $V_{\rm eff}$ (circles with error bars) and $V_{\rm max}$ (lines). Except for the slight difference in the sample of $\mhi>10^8\msun$, which is within the measurement errors, there is no significant difference of using the two volume weights for higher \hi\ mass samples.

\subsubsection{NGC versus SGC}
\begin{figure*}
	\centering
	\includegraphics[width=\textwidth]{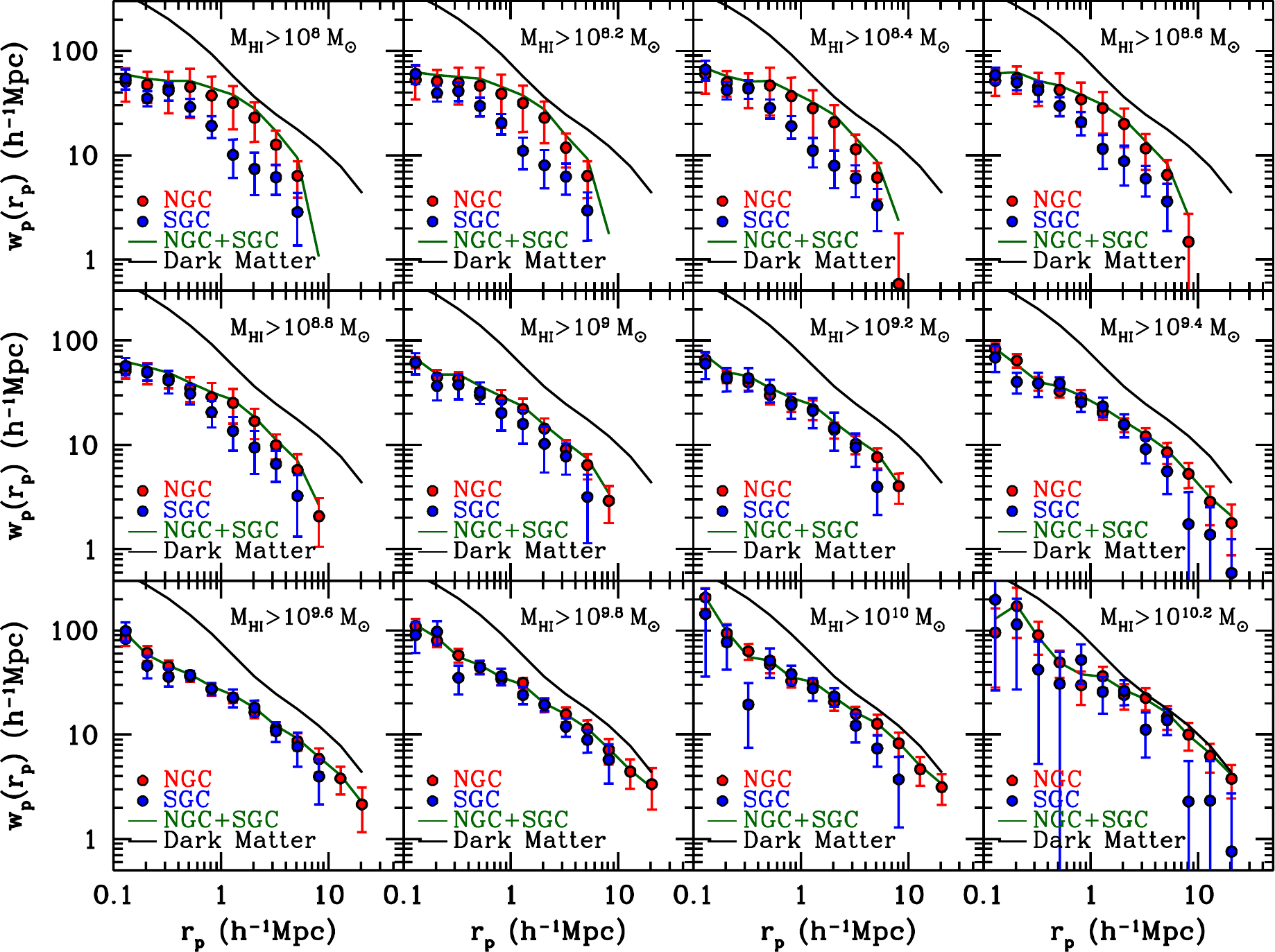}
	\caption{Projected 2PCF measurements for galaxies in the northern (red filled circles) and southern galactic caps (blue filled circles). The different panels show the measurements from different \hi-mass threshold samples. The green line in each panel displays the results from the combined sample of the NGC and SGC, while the black line shows the projected 2PCF of dark matter as in Figure~\ref{fig:wp_MHI}.
	} \label{fig:nssplit}
\end{figure*}
Another straightforward way to investigate the effect of sample variance and the validity of our results is to compare the clustering measurements from the data in the northern galactic cap (NGC) and the southern galactic cap (SGC). The Virgo and Coma superclusters are both located in the northern galactic hemisphere, while the Perseus-Pisces supercluster, which dominates the southern hemisphere, is mostly outside the survey area of the $\alpha$.70 sample. The results are shown in Figure~\ref{fig:nssplit}, where we plot the measurements of $w_{\rm p}(r_{\rm p})$ for the same \hi\ mass thresholds, but separately for the NGC (red filled circles) and SGC (blue filled circles) data. The measurements with the full data for each mass threshold sample and the projected 2PCF of dark matter are also shown as green and black solid lines for comparison. 

The $\wprp$ measurements with the NGC and SGC data generally agree with each other within $1\sigma$ errors for high mass samples with \hi\ mass thresholds above $10^9\msun$, indicating that the local superclusters are no longer a dominant source of uncertainty when \hi\ mass exceeds this threshold. When galaxies with $\mhi<10^9\msun$ are included, however, we find the NGC samples have stronger clustering at intermediate scales than the corresponding SGC samples, and the effect is stronger for lower masses. At the lowest mass with $\mhi>10^8\msun$, the $\wrp$ at $\rp\sim 2\mpchi$ obtained from the NGC sample is about four times that obtained from the SGC sample.  The figure also reveals that the NGC sample dominates the contribution to the measurements of the full sample at each given mass threshold. If only the SGC is considered for all masses, one would observe very weak or no mass dependence of the clustering for mass thresholds below $\sim10^{9}\msun$, instead of the declining trend seen in the total and NGC samples. 

This analysis confirms our conjecture that the enhancement of the clustering at intermediate scales as seen in Figure~\ref{fig:wp_MHI} is contributed mostly (if not completely) by the galaxies in the NGC, where the several known superclusters dominate the clustering of the low-mass samples. Although the NGC sample covers a volume that is about 2.7 times larger than the SGC sample, the $\wrp$ errors of the former at low \hi-mass are comparable to, or even larger than the latter. This again reflects the dominating role of the local superclusters in the clustering of the NGC galaxies. However, as seen from the redshift distribution of galaxies in the SGC (magenta histogram) in Figure~\ref{fig:selection}, there are also clear imprints of voids and clusters, but not as significant as in the NGC. These voids could potentially lower the measured clustering amplitudes. Since the effect of the local large-scale structures is already taken into account in the $V_{\rm eff}$ weight, we expect the $\wprp$ measurements from the SGC to be more reliable for the low \hi-mass samples.

\subsubsection{Clustering relative to an SDSS reference sample}
So far our results of \hi-mass dependent clustering are presented with the auto-correlation functions, which suffer from sample variance effects given the limited survey volume. The other way to study the \hi-mass dependence of the clustering with an alleviated sample variance effect is to cross-correlate each \hi\ sample (target sample) with a reference sample. The reference samples for all target samples are selected so that galaxies in them are controlled to have the same distribution in intrinsic properties, but galaxies in each reference sample are weighted to give a similar redshift distribution as the corresponding target sample. 
We can derive the ratio of bias factors of the target sample and the corresponding reference sample from the target-reference and reference-reference correlation functions measured within the ALFALFA footprint. This ratio is largely free of sample variance. To obtain the bias factor of each target sample from this ratio, we can use the bias factor inferred from the reference-reference correlation function measured in a much larger volume (within the SDSS footprint, see below). In this step, the sample variance from the clustering of the reference sample still exists, but with a smaller magnitude as a result of the larger volume. Overall, with the above procedure, we can achieve measurements of bias factors for the target \hi\ galaxy samples with a smaller sample variance effect.

We apply the cross-correlation technique as follows. First, we construct a volume-limited tracer sample in the SDSS Main galaxy sample, which consists of 27,411 galaxies with $r$-band absolute magnitudes $-19.5<M_r<-18.5$ and redshifts $0.0025<z<0.05$. Next, a {\em reference} sample is selected from the above tracer sample by applying the geometry of the ALFALFA survey. For a given \hi-selected sample, we assign a weight of $n_{\rm HI}(z)/n_{\rm reference}(z)$ to each galaxy in the {\em reference} sample to ensure the same redshift distribution as in the \hi\ sample. Finally, the counterpart of each reference sample in the full SDSS footprint is also similarly constructed. As the SDSS Main sample covers only three narrow stripes in the SGC with a relatively small sky coverage, we concentrate on the NGC for both the ALFALFA target and SDSS reference samples when measuring the cross-correlation functions.

The cross-correlation function between a given \hi-selected sample and the corresponding reference sample is estimated using the extended Landy-Szalay estimator as in \citet{Zehavi02} and \citet{Guo12},
\begin{equation}
\xi(r_{\rm p},r_{\rm\pi})=\frac{\rm{D_1D_2-D_1R_2-D_2R_1+R_1R_2}}{\rm{R_1R_2}},
\end{equation}
where the subscripts ``1'' and ``2'' stand for the \hi-selected sample and the reference sample, and $\rm{D}$ and $\rm{R}$ for real and random samples, respectively. The $1/V_{\rm max}$ weight is similarly applied as in Eqs.~(\ref{eq:dd})--(\ref{eq:rr}) and the galaxy weights in the reference sample are accounted for in doing the pair count. Similarly we measure the 2PCF of the reference sample. For both the target-reference and reference-reference correlation functions, we integrate $\xi(r_{\rm p},r_{\rm\pi})$ along the line of sight up to $r_{\pi,{\rm max}}=20\mpchi$ to obtain the projected 2PCFs. The relative bias, $b/b_{\rm REF}$, between a given \hi\ sample and the reference sample is expected to follow
\begin{equation}
w_{\rm p,HIxREF}=(b/b_{\rm REF})w_{\rm p,REFxREF}, \label{eq:bref}
\end{equation}
where $w_{\rm p,HIxREF}$ and $w_{\rm p,REFxREF}$ denote the projected target-reference and reference-reference two-point correlation functions, respectively.
To determine the relative bias from the data, we use the $w_{\rm p,HIxREF}$ and $w_{\rm p,REFxREF}$ measurements at $r_{\rm p}>5\mpchi$ and find the best-fitting value of $b/b_{\rm REF}$ by maximizing the likelihood function,
\begin{equation}
    \mathcal{L} \propto \frac{1}{\sqrt{\mathbf{|C|}}}\exp(-\chi^2/2),
\end{equation}
where
\begin{equation}
\chi^2 = (\mathbf{w_{\rm p,HIxREF}}-\mathbf{w_{\rm p,HIxREF}^*})^T
\mathbf{C}^{-1}(\mathbf{w_{\rm p,HIxREF}}-\mathbf{w_{\rm p,HIxREF}^*}). \label{eq:chi2}
\end{equation}
The quantity with (without) a superscript `$*$' is the one from the model of equation~(\ref{eq:bref}) (data). The covariance matrix depends on $q=b/b_{\rm REF}$ as,
\begin{equation}
\mathbf{C} = \mathbf{C}_{\rm HIxREF}-q(\mathbf{C}_{\rm HI,REF}+\mathbf{C}^T_{\rm HI,REF})+q^2\mathbf{C}_{\rm REFxREF},
\end{equation}
where $\mathbf{C}_{\rm HIxREF}$ is the covariance matrix of $w_{\rm p,HIxREF}$, $\mathbf{C}_{\rm REFxREF}$ that of $w_{\rm p,REFxREF}$, and    $\mathbf{C}_{\rm HI,REF}$ the covariance between $w_{\rm p,HIxREF}$ and $w_{\rm p,REFxREF}$. 

We also measure the projected 2PCF of the counterpart of each reference sample constructed within the SDSS footprint and infer $b_{\rm REF}$ by comparing to the projected 2PCF of matter. With the ratio $b/b_{\rm REF}$ and $b_{\rm REF}$, we end up with an inference of the bias factor $b$ for each \hi\ galaxy sample. The results are shown as squares in the left panel of Figure~\ref{fig:bias}. The mass dependence is clear for galaxies with $\mhi>10^9\msun$, while at lower \hi\ mass the dependence disappears but with large error bars. The result will be discussed later with bias factors inferred from other methods. 

\begin{figure}
	\centering
	\includegraphics[width=0.4\textwidth]{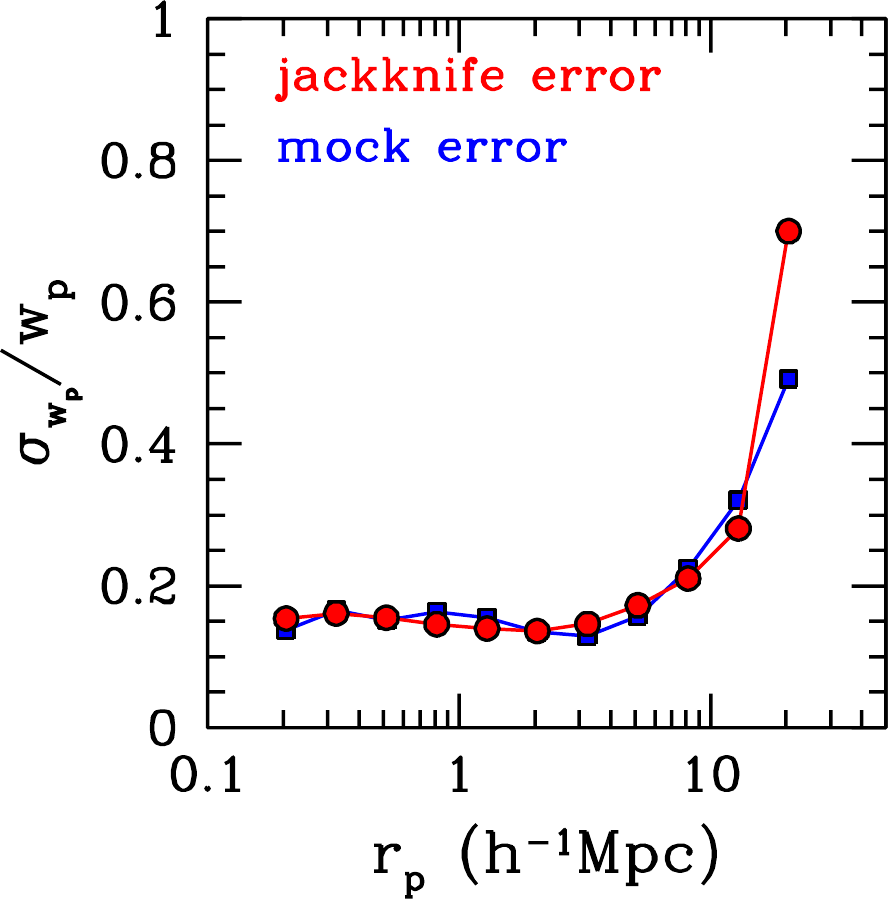}
	\caption{Comparison between the average fractional errors of the projected 2PCFs using the jackknife resampling method (red circles) and those estimated from mock galaxy catalogs (blue squares) constructed from sub-volumes of a large simulation box for $\mhi>10^{10}\msun$.} \label{fig:jack}
\end{figure}

\subsubsection{Comparison with mock catalogs}
\label{sec:mocks}

The effect of sample variance can be evaluated accurately using $N$-body simulations with constrained initial conditions to simulate the dark matter distribution in the local universe as proposed in \cite{Zavala09}. For the current work we have constructed a set of 64 mock galaxy catalogs which have the same geometry as the ALFALFA sample (see detailed description in the next section). We do not attempt to obtain and apply constrained initial conditions to our simulation, which needs substantially more work and is not necessary for our purpose. In order to test whether the apparent clustering dependence on the \hi\ mass at the high-mass end could be weakened by potential underestimates of the measurement errors from the jackknife resampling, we select 64 different sub-volumes of the SMDPL simulation box and randomly place a virtual observer in each of them. We construct a mock catalog from each sub-volume by mimicking the geometry of the ALFALFA survey, and measure $\wprp$ in the same way as above. The correlation function errors caused by sample variance are then estimated from the variations among these mock catalogs. Figure~\ref{fig:jack} compares the average fractional diagonal errors of the projected 2PCFs as estimated from the jackknife resampling method (red circles) and those from the mock catalogs (blue squares) for the $\mhi>10^{10}\msun$ sample. The estimates from the two methods agree well with each other at all scales except the largest scale probed ($\rp\sim20\mpchi$), where the jackknife error is larger than the mock error by about 40\%. We also find the same result when using other \hi\ mass samples. This is in good agreement with previous studies (see Appendix B of \citealt{Guo13} for more details). This demonstrates that the jackknife errors are accurate on most scales probed here, from $\sim20\kpchi$ up to about $20\mpchi$, but more conservative at larger scales. 

The finite sample volume may cause not only the variance from measurement to measurement, but also a systematic underestimate of the 2PCF by a constant value, known as the ``integral constraint''. We use the mock catalogs constructed above to quantify this effect, by comparing the $w_{\rm p}(r_{\rm p})$ averaged from different mocks with the one measured from the whole simulation box. The difference between the two $w_{\rm p}$ measurements on large scales is a measure of the integral constraint, which is $\sim0.6\mpchi$ for  the sample with $\mhi>10^9\msun$. The difference increases to $\sim1.5\mpchi$ if only the SGC is considered, explaining the slightly larger clustering amplitudes from the full-area sample than from the NGC alone, as shown in Figure~\ref{fig:nssplit}. For samples with lower threshold masses, the effective volume drops and the bias factor decreases, which have opposite effects on the magnitude of the integral constraint. We find that the integral constraint shows a mild increase toward low mass samples, and the data point at $\sim 8\mpchi$ in each top panel of Figure~\ref{fig:wp_MHI} is the only one substantially affected by the integral constraint (the correction is estimated to be at the level of $\Delta w_{\rm p}=2.5\mpchi$ for the low \hi\ mass samples from mock tests described in the next section). The differences seen between NGC and SGC measurements (Figure~\ref{fig:nssplit}) are mainly caused by sample variance. Therefore, the \hi\ mass dependence of the clustering on large scales and at high masses as observed above is robust.

\subsection{Redshift-space monopole moments and the biasing of the \hi-selected galaxies}

\begin{figure*}
	\centering
	\includegraphics[width=\textwidth]{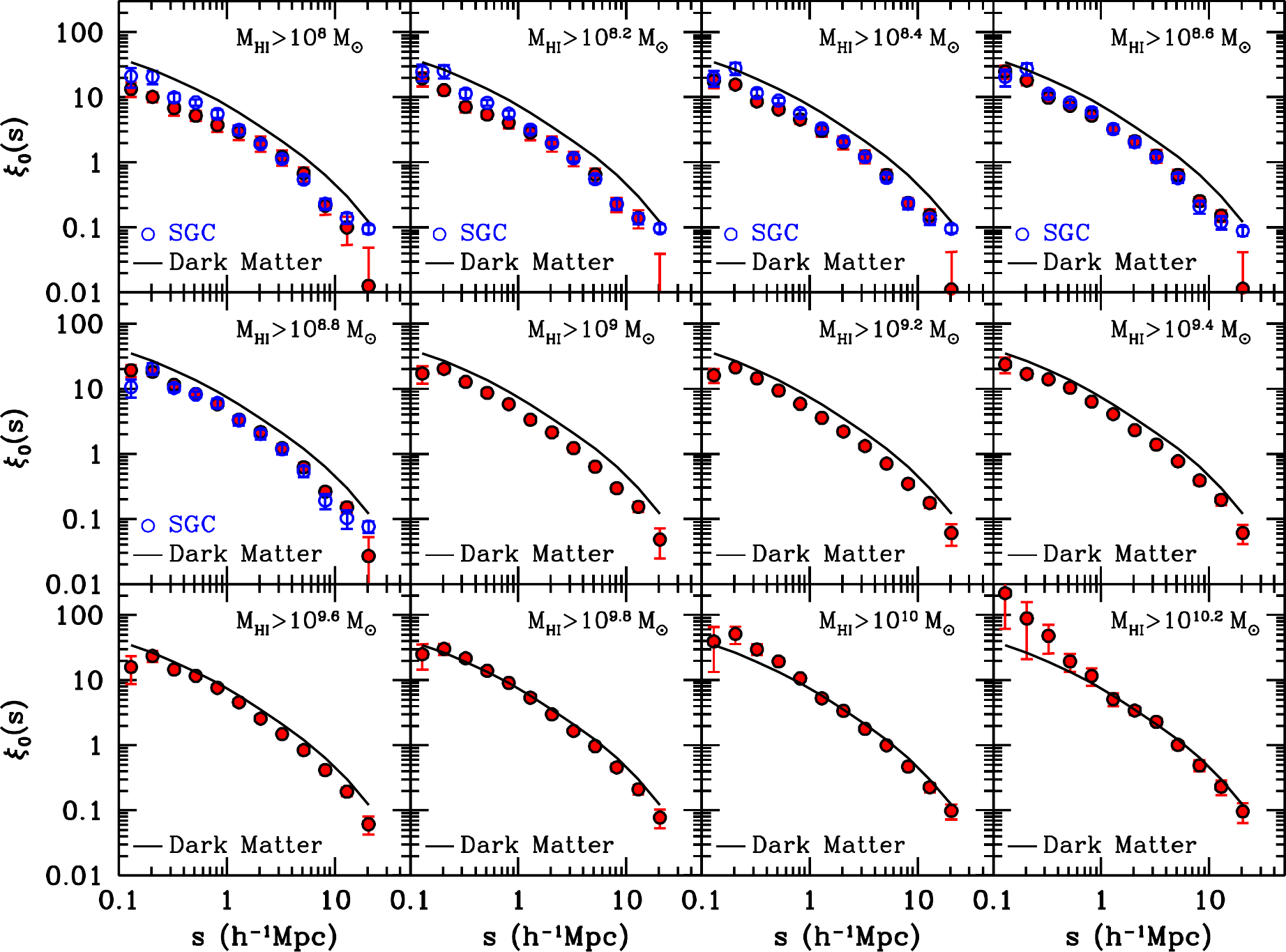}
	\caption{Similar to Figure~\ref{fig:wp_MHI}, but for the redshift-space monopole moment $\xi_0(s)$. The filled red circles represent the measurements of $\xi_0(s)$ for the different $\mhi$-threshold samples using the combined sample of the NGC and SGC, while the open blue circles for the low \hi-mass samples are measured from the SGC. The black solid line in each panel is the corresponding monopole moment predicted for the linear fluctuation field of dark matter (i.e., with a bias factor of $b=1$).
	} \label{fig:xi_MHI}
\end{figure*}

\begin{figure*}
	\centering
	\includegraphics[width=0.8\textwidth]{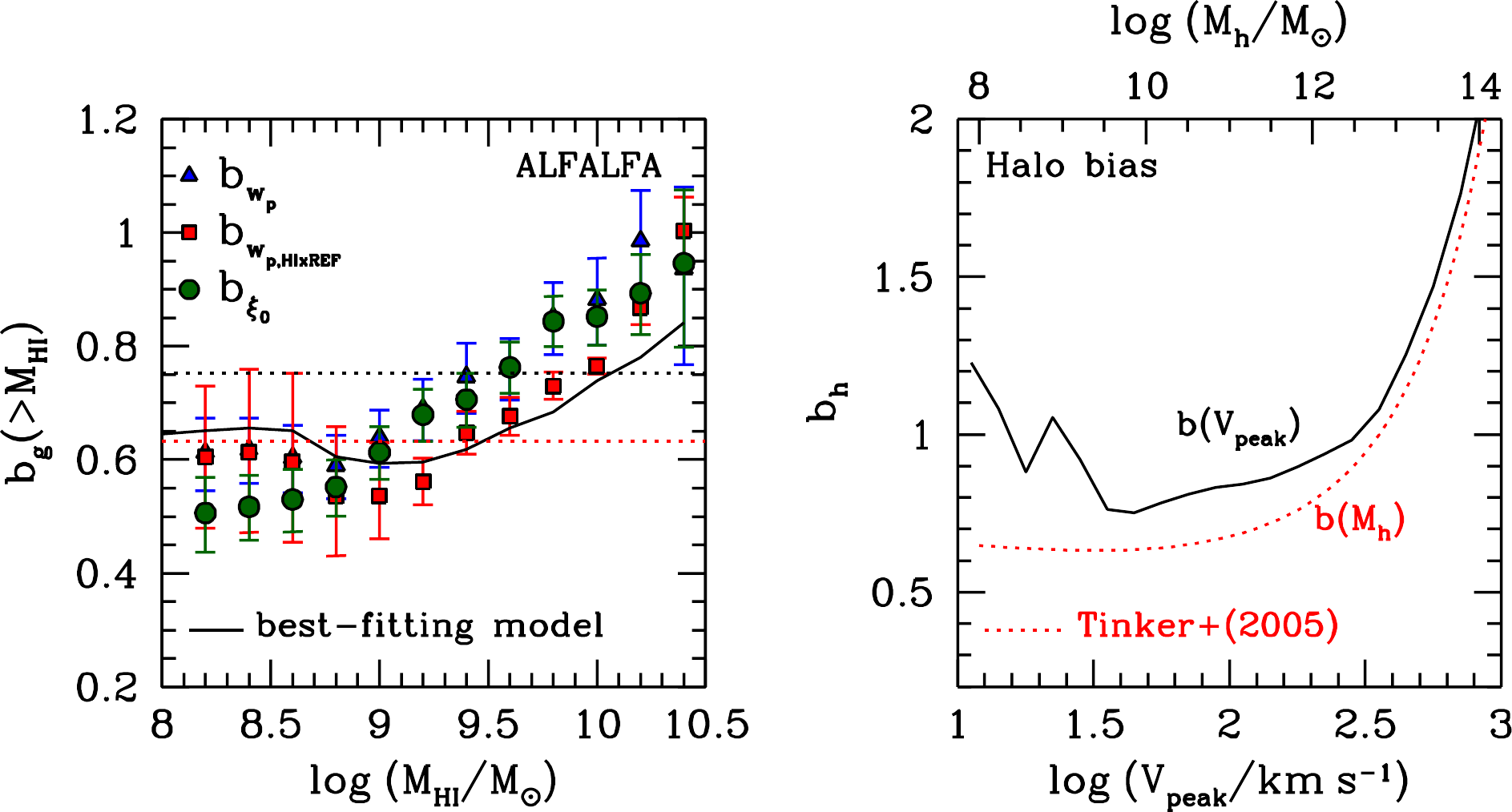}
	\caption{Left: galaxy bias factor $b_{\rm g}({>}\mhi)$ inferred from different methods for different $\mhi$-threshold samples. The methods include galaxy-to-matter $w_{\rm p}$ ratio (triangles), cross-correlation with an SDSS reference sample (squares),   redshift-space monopole moment [Equation~(\ref{eq:bias})] (circles), and result from an extended SHAM model fit (solid curve). See the text for details. The red (black) horizontal dotted line represent the minimum halo bias factor from the mass ($V_{\rm peak}$) dependent dark matter halo clustering. Right: halo bias factor $b_{\rm h}$ as a function of halo mass (dotted; \citealt{Tinker05}) or $V_{\rm peak}$ (solid). 
	} \label{fig:bias}
\end{figure*}

In addition to the measurements of projected 2PCFs as presented above, we have also measured the monopole moment $\xi_0(s)$ of the redshift-space three-dimensional (3D) 2PCF for our \hi-selected samples, from which we aim to estimate the galaxy bias factor as a function of \hi\ mass. In principle, $\wprp$ is preferred over $\xi_0(s)$ for the purpose of measuring galaxy bias, as the redshift-space distortion (RSD) is expected to be minimized in the former. In our case, however, we have adopted a relatively small integration depth with $r_{\pi,{\rm max}}=20\mpchi$ in order to achieve the best S/N. This may result in a significant residual RSD effect on our $\wprp$ measurements at large scales, leading to biased determination of the galaxy bias factors. Therefore, for the purpose of measuring the galaxy bias factors we opt for $\xi_0(s)$ which is not affected by the choice of $r_{\pi, {\rm max}}$. 

Following common practice we first estimate the redshift-space 3D 2PCF in the form of $\xi(s,\mu)$ for each of our \hi-mass threshold samples, where $\mathbf{s}$ is the 3D separation in redshift space with the amplitude defined by $s^2=r_{\rm p}^2+r_{\pi}^2$ and $\mu$ is the cosine of the angle between $\mathbf{s}$ and the line of sight. The redshift-space monopole moment is then obtained by integrating $\xi(s,\mu)$ over the full range of $\mu$, 
\begin{equation}
\xi_0(s)=\int_{0}^{1}\xi(s,\mu){\rm d}\mu=\sum_i\xi(s,\mu_i)\Delta\mu .
\end{equation}
We adopt linear bins for $\mu$ ranging from 0 to 1 with a constant interval of $\Delta\mu=0.05$. On linear scales, the redshift-space monopole moment is related to the real-space 2PCF $\xi(r)$ through the following equation \citep{Kaiser87,Hamilton92},
\begin{eqnarray}
\xi_0(s)&=&\left(1+\frac{2}{3}\beta+\frac{1}{5}\beta^2\right)\xi(r)\nonumber\\
&=&\left(1+\frac{2}{3}\beta+\frac{1}{5}\beta^2\right)b_{\rm g}^2\xi_{\rm DM}(r), \label{eq:bias}
\end{eqnarray}
where $\beta=f/b_g$ and $\xi_{\rm DM}$ is the real-space 2PCF of dark matter. The linear growth rate $f\simeq\Omega_{\rm m}^{0.55}$ at $z=0$ for our adopted cosmology. In practice, we use the measurements of $\xi_0(s)$ at $s>5\mpchi$ to determine the value of $A\equiv b_{\rm g}^2(1+2\beta/3+\beta^2/5)$ by minimizing the $\chi^2$,
\begin{equation}
\chi^2= (\bm{\xi}_0-A\bm{\xi}_{\rm DM})^T \mathbf{C}^{-1}(\bm{\xi}_0-A\bm{\xi}_{\rm DM}), 
\end{equation}
where $\mathbf{C}$ is the full error covariance matrix of large-scale measurements of $\xi_0(s)$. The errors on $b_{\rm g}$ are estimated from the probability distribution with $\Delta\chi^2=1$. In order to reliably measure the galaxy bias, we have also corrected for the integral constraint effect by using the mock galaxy catalogs (see details in \S\ref{sec:sham}).

Figure~\ref{fig:xi_MHI} displays the measurements of $\xi_0(s)$ for the different $\mhi$-threshold samples. The corresponding monopole for dark matter distribution is plotted as a solid black line and repeated in every panel for comparison. For samples with mass thresholds below $10^9\msun$, we additionally show the result from the SGC as blue circles. At a given scale the monopole moment increases with increasing mass threshold, and the effect is similarly seen at all scales. This result is broadly consistent with the \hi\ mass dependence of the projected 2PCFs, demonstrating that the mass dependence observed from the $\wprp$ measurements is not an artifact of the residual RSD effect. 

It is interesting that the SGC and the full area agrees very well in the monopole moment measurement at all scales except the largest scales, where the $\xi_0(s)$ from the SGC becomes more noisy and drops quickly as a result of the limited sample volume. This finding suggests that, for small-volume samples like our \hi-selected galaxy samples, the monopole moment is less affected by the sample variance when compared to the projected 2PCF. As discussed above, the sample variance effect in $w_{\rm p}(r_{\rm p})$, manifested as the enhancement at intermediate scales in Figure~\ref{fig:nssplit}, is dominated by the Virgo Supercluster in the NGC, which effectively boosts the number of galaxy-galaxy pairs but with a broadened distribution of the line-of-Sight pair separation.  It thus enhances the line-of-sight integration of $\xi(r_{\rm p},r_{\rm\pi})$, leading to higher $\wrp$ even at small $\rp$. In the case of the monopole moment, the galaxy pairs in Virgo with large line-of-sight separations contribute exclusively to the large-scale measurement, with little effect on the intermediate-to-small scales. The monopole moments thus provide an additional robustness test on our finding of the \hi-mass dependence of galaxy clustering, which was obtained above from the measurements of $\wprp$.

The galaxy bias factor $b_{\rm g}({>}\mhi)$ inferred from Eq.~(\ref{eq:bias}) is plotted as circles in the left panel of Figure~\ref{fig:bias} as a function of \hi\ mass threshold. The bias factor increases nearly linearly with increasing $\mhi$ threshold, ranging from $\sim0.5$ at the lowest mass ($\sim10^{8.2}\msun$) up to $\sim0.9$ at the highest mass ($\sim10^{10.2}\msun$). Also shown in the panel are the bias factors estimated based on the galaxy to matter $w_{\rm p}$ ratio (triangles) and the cross-correlation technique (squares). The solid curve corresponds to the best-fitting theoretical model (\S~\ref{sec:sham}). The ones from cross-correlation and best-fitting model are lower than other estimates above $\mhi>10^9\msun$. At the low mass end, the large error bars prevent us from seeing a clear trend. Overall the different inferences appear to be consistent, given the error bars. 

For comparison, we show in the right panel of Figure~\ref{fig:bias} the bias values $b_{\rm h}$ for dark matter halos. The dotted curve shows $b_{\rm h}$ as a function of halo mass from the fitting formula of \cite{Tinker05}, and the solid curve shows $b_{\rm h}$ as a function of $V_{\rm peak}$ from the SMDPL simulation. The halo mass dependent bias factor approaches a plateau of $\sim 0.63$ at low mass, and the $V_{\rm peak}$ dependent bias factor reaches a minimum of $\sim 0.75$ around tens of $\rm km\, s^{-1}$. These two bias values are marked with the red and black dotted lines in the left panel. The galaxy samples with low \hi\ masses show a hint of bias factors lower than expected from low mass or low $V_{\rm peak}$ halos. It indicates that it would be hard to interpret the clustering of \hi\ galaxies with simple halo-based models. If the result persists with better data from future surveys, it would be remarkable. In \S~\ref{sec:model}, we will discuss the implications.

\subsection{Section summary}

Before going to the next section for theoretical modeling, we briefly summarize our observational results. We have estimated the projected 2PCF $\wprp$, the projected cross-correlation with a reference sample, and the monopole moment $\xi_0(s)$ of the redshift-space 2PCF, for the 13 $\mhi$-threshold samples. The three types of measurements can all be used to estimate galaxy bias factors. While they suffer from different systematics  (e.g., residual RSD effect with the projected correlation functions and inaccuracy of Kaiser formula in the weakly non-linear regime with the monopole method), the inferred galaxy bias factors are in broad agreement. We find that the galaxy bias factor increases nearly linearly with increasing threshold $\mhi$. There is also a hint that galaxy bias factors for low \hi\ mass samples are lower than expected from low mass (or low $V_{\rm peak}$) halos. In the next section, we perform modeling based on the $\wrp$ measurements.  At low \hi\ masses those measurements are affected by sample variance, but as we have demonstrated, the effect is properly included in the error estimation. This results in relatively large errors in $\wprp$ measurements for low-mass samples, limiting their constraining power on the theoretical models. For this reason one may want to drop the low-mass samples or the galaxies in the NGC when doing the modeling. However, as pointed out by \cite{Norberg11}, any non-Bayesian massaging of the data is not recommended in clustering analyses, unless one knows the true answer. In fact, as we will show, using only the SGC data would not significantly change our model parameters.

\section{Modeling the clustering of \hi-selected galaxies} \label{sec:model}

With the $\wprp$ measurements of the $\mhi$ threshold samples, we perform theoretical modeling to investigate the connection between \hi\ galaxies and dark matter halos. We first discuss the difficulty with the commonly adopted form of the HOD model to explain the measurements, and show that the subhalo abundance matching (SHAM) model in its simplest form is unable to reproduce the measurements, either. We then extend the SHAM model by including additional parameters related to the assembly history of halos/subhalos. Finally, we present the modeling results based on an extended SHAM model that incorporates halo/subhalo formation time.

\subsection{Modeling with the HOD model}\label{subsec:hod}

Statistical models of galaxy distribution have been developed to link galaxies of different properties (e.g., luminosity, color, stellar mass, and star formation rate) with dark matter halos. These include the halo occupation distribution model \citep[HOD; e.g.,][]{Jing98,Peacock00,Seljak00,Scoccimarro01,Berlind02,Zheng05,Zheng09,Leauthaud12,Geach12,Guo14b,Guo15b,Skibba15,Zu15}, the conditional luminosity function model \citep[CLF; e.g.,][]{Yang03,Yang04,Yang12,vandenBosch13}, the subhalo abundance matching model \citep[SHAM; e.g.,][]{Kravtsov04,Vale-Ostriker-04,Conroy06,Shankar-06,Vale06,WangL07,Behroozi10,Guo10,Moster10,Nuza13,Rodriguez-Puebla13,Sawala15}, and halo-based empirical model \citep[e.g.][]{LuZ14,LuZ15}.

We first make an attempt to model the measurements with the HOD model. For a threshold galaxy sample, the central galaxy occupation is commonly parametrized as \citep{Zheng05,Zheng07},
\begin{equation}
\langle N_{\rm cen}(M_{\rm h})\rangle=\frac{1}{2}\left[1+{\rm erf}\left(\frac{\log M_{\rm h}-\log M_{\rm min}}{\sigma_{\log
M}}\right)\right], \label{eq:Ncen}
\end{equation}
where ${\rm erf}$ is the error function, $M_{\rm min}$ is the cutoff halo mass of central galaxies at which the occupation number $\langle N_{\rm cen}(M_{\rm min})\rangle=0.5$, and $\sigma_{\log M}$ characterizes the width of the cutoff profile. The mean occupation function of satellites follows a power law modified by a low mass cutoff \citep{Zheng05,Zheng07},
\begin{equation}
\langle N_{\rm sat}(M_{\rm h})\rangle=\langle N_{\rm cen}(M_{\rm h})\rangle
\left(\frac{M-M_0}{M_1^\prime}\right)^\alpha,
\label{eq:Nsat}
\end{equation}
and the number of satellites at fixed halo mass is assumed to follow a Poisson distribution. This five-parameter ($M_{\rm min}$, $\sigma_{\log M}$, $M_0$, $M_1^\prime$, and $\alpha$) model works well for luminosity-threshold or stellar-mass-threshold samples \citep[e.g.,][]{Zehavi11}.

When we apply the above HOD form to model the $w_{\rm p}$ measurements of the \hi\ galaxies, we find that we could not achieve good fits, with the model $w_{\rm p}$ having too high amplitudes. The main reason lies in the tension between galaxy number density and clustering amplitude (bias factor). For each galaxy sample, the number density of galaxies roughly determines a halo mass threshold, and halos above such a threshold appear to have higher clustering amplitude than galaxies. 

One way to reduce the tension is to introduce a duty cycle parameter $f_{\rm dc}$ (i.e., by multiplying this parameter with the above mean occupation function), which can account for the fact that only a fraction of halos at fixed mass can host the observed \hi\ galaxies. With the duty cycle parameter, we can populate \hi\ galaxies into lower mass halos (with lower bias factor) while still matching the galaxy number density. Indeed, we are able to obtain reasonable fits. However, the fitting results for low $\mhi$ samples appear to be unphysical. For example, the value of $M_{\rm min}$ for the $\mhi>10^8\msun$ sample is $1.3\times 10^9\msun$. That is, about half of the baryons associated with halos of $M_{\rm min}$, in the sense of applying a global baryon fraction, is in the form of \hi\ gas. For the $\mhi>10^9\msun$ sample, the result is even more absurd, with  $M_{\rm min}=1.2\times 10^9\msun$ (i.e., a $M_{\rm min}$ halo would be almost wholly made of \hi\ gas).

Given the difficulty in interpreting the clustering measurements of \hi\ galaxies with a simple HOD parametrization, we turn to the SHAM model.

\subsection{Modeling with the SHAM model}
\label{subsec:sham}
\begin{figure*}
	\centering
	\includegraphics[width=0.9\textwidth]{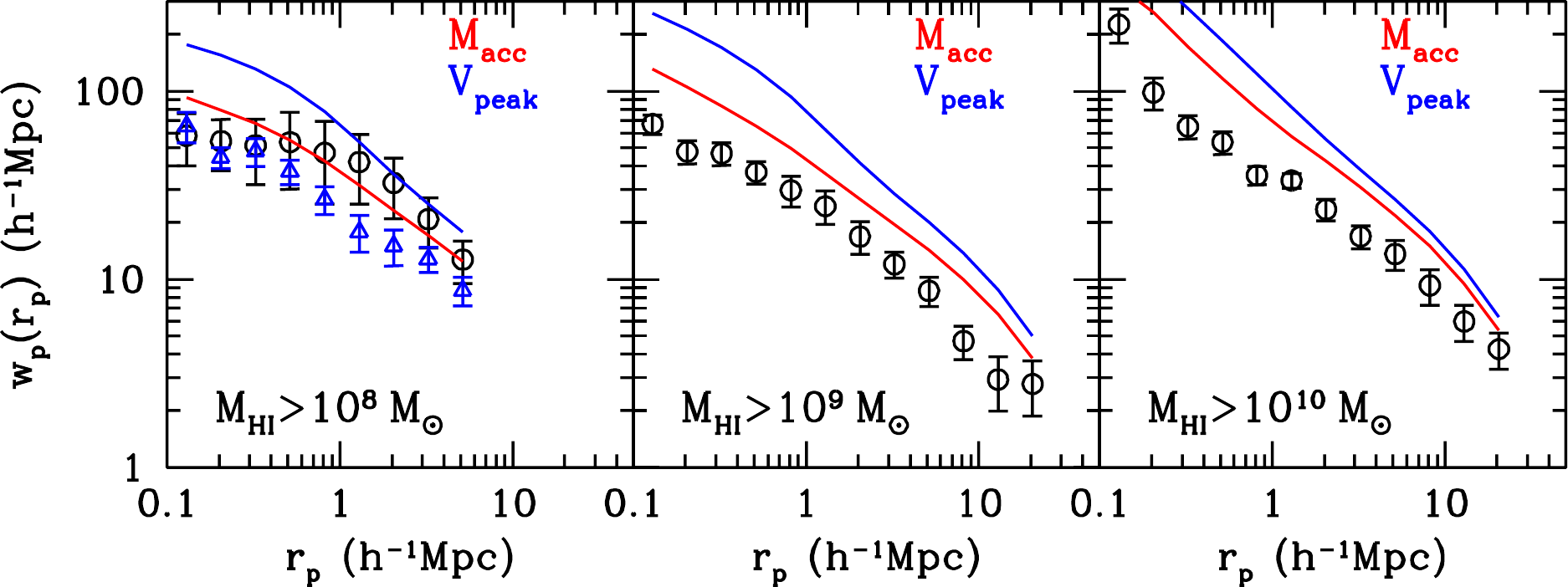}
	\caption{Projected 2PCFs from two SHAM models with $V_{\rm peak}$ (blue lines) and $M_{\rm acc}$ (red lines) as the halo properties for three \hi-selected galaxy samples. The symbols with errors are the data measurements of ALFALFA samples. Blue triangles in the left panel represent measurements from the SGC data alone.
	} \label{fig:shamcomp}
\end{figure*} 
With the SHAM model, we adopt the simplest form as our starting point. The SHAM model assumes that the stellar mass (or luminosity) of a galaxy is an increasing function of the maximum mass or circular velocity ever attained by its halo and derives the relation between stellar mass and halo mass by matching the abundance of galaxies above a given stellar mass threshold with that of halos/subhalos above a threshold in maximum halo mass or circular velocity. The result agrees well with the stellar-to-halo mass relation established for central galaxies in groups/clusters in the low-redshift universe \citep[see e.g.,][]{Conroy09,Behroozi10,Guo10,Reddick13}. Recently the subhalo age distribution matching (SADM) is proposed as an extension to SHAM to further model galaxy color at fixed stellar mass, assuming redder galaxies are hosted by halos/subhalos that are formed earlier \citep{Hearin-Watson-13}. These simple models can reasonably reproduce the dependence of clustering on stellar mass (or luminosity) and color (see Fig.~3 of \citealt{Hearin-Watson-13} for an example), suggesting a possible connection between galaxy assembly and halo assembly. 

We start with the SHAM model, attempting to populate the dark matter halos in the SMDPL simulation with galaxies of different \hi\ masses. We consider two halo parameters: the peak circular velocity ($V_{\rm peak}$) over the entire merger history of the (sub)halo, and the halo mass. Following common practice, we use the current halo mass for a main halo supposed to host a central galaxy, and the mass at the last epoch of accretion for subhalos supposed to host satellite galaxies, both referred as $M_{\rm acc}$ in what follows. We rank-order all the halos/subhalos in the SMDPL simulation by either $V_{\rm peak}$ or $M_{\rm acc}$, and populate halos/subhalos of larger $V_{\rm peak}$ or $M_{\rm acc}$ with galaxies of higher $\mhi$. In Figure~\ref{fig:shamcomp}, the $\wprp$ predicted by this simple model is compared with the observational measurements for three \hi\ mass thresholds. Results from the $V_{\rm peak}$- and $M_{\rm acc}$-based models are shown as blue and red curves, respectively. Both models predict too strong clustering on all scales and at all \hi\ mass thresholds, except the lowest mass threshold ($\mhi>10^8\msun$) for which the $M_{\rm acc}$-based model roughly matches the data. However, this agreement should not be overemphasized: as discussed in the previous section, the enhanced clustering at intermediate scales as seen in the low-mass samples is likely a biased measurement caused by sample variance (see the measurements with the SGC data). 

In the simplest form of the SHAM model we adopt, no scatter is introduced between galaxy \hi\ mass and halo/subhalo mass (or $V_{\rm peak}$). The failure of the model for the high $\mhi$ samples can be alleviated by introducing such a scatter, which allows a faction of \hi\ galaxies to be populated into lower mass/$V_{\rm peak}$ halos (which have lower halo bias and weaker clustering). We find, however, that a simple inclusion of scatter up to 2 dex in stellar mass does not help for high $\mhi$ samples, presumably caused by the tension between the requirements of halo mass/$V_{\rm peak}$ thresholds from galaxy number density and low galaxy clustering amplitude. Scatter does neither help for low $\mhi$ samples, as the large-scale halo bias factor approaches a plateau/minimum toward low mass/$V_{\rm peak}$ ($b_{\rm h}(M_{\rm h})\sim0.63$ at halo mass below $\sim 10^{11}\msun$ and $b_{\rm h}(V_{\rm peak})$ has a minimum of 0.75 as shown in Figure~\ref{fig:bias}). As the galaxy bias factor $b_g({>}M_{\rm HI})$ tends to go below such values for low \hi\ mass threshold samples (left panel of Figure~\ref{fig:bias}), the SHAM model with scatter cannot explain the measured clustering for low $\mhi$ threshold samples. This remains true for any model that assigns \hi\ galaxies based only on the mass of dark matter halos, even if those galaxies are populated into a randomly selected sub-population of halos. This implies that halos of similar dark matter masses are not randomly populated by \hi-selected galaxies. Therefore, to properly model the \hi-rich galaxies, one needs to introduce additional halo parameters that are related to the cold gas supply of galaxies, and those associated with the assembly history of halos are natural candidates.

\subsection{Extending the SHAM model by introducing additional halo parameters}

\begin{figure*}
	\centering
	\includegraphics[width=0.8\textwidth]{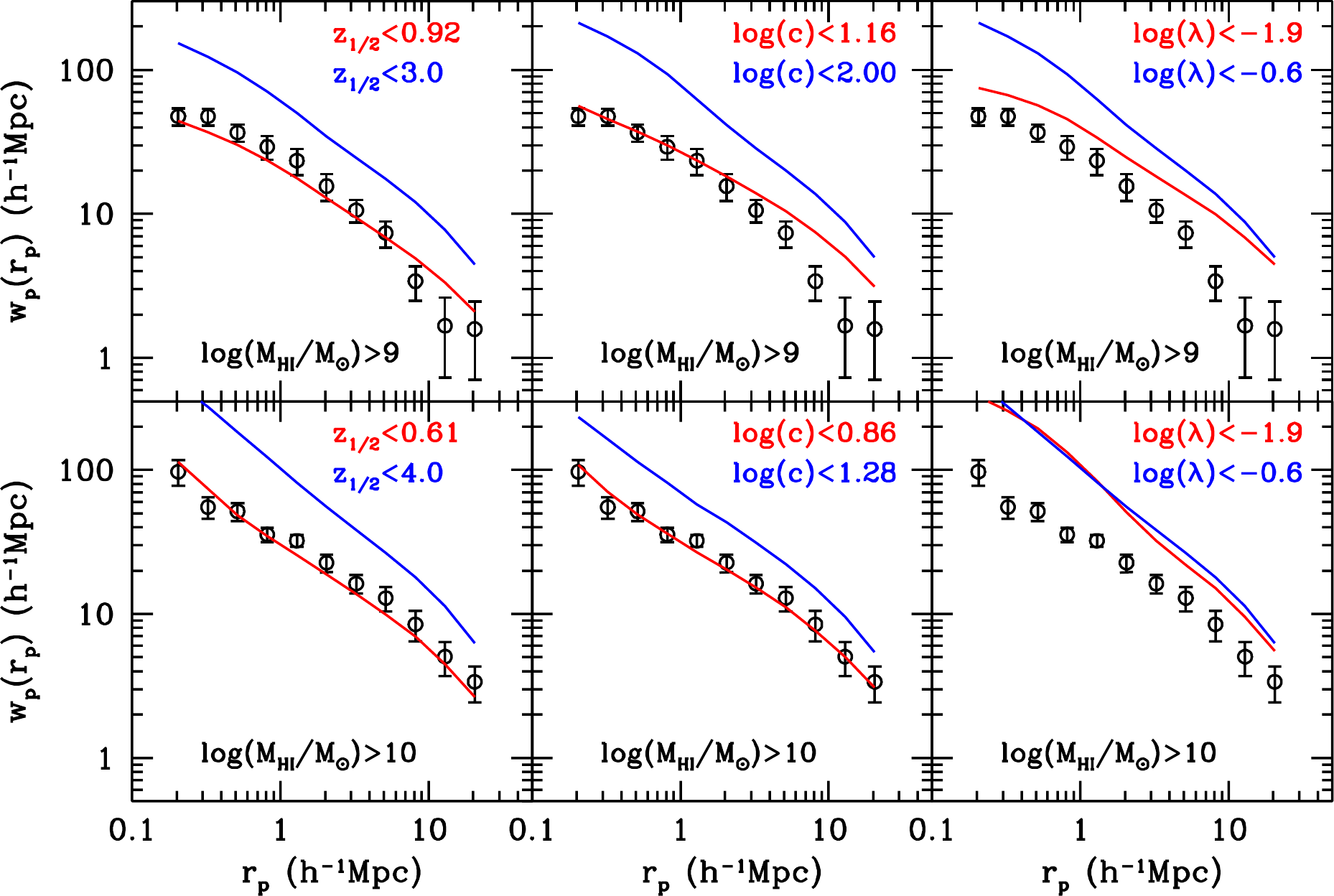}
	\caption{Models taking into account halo assembly parameter for the samples of $\mhi>10^9\msun$ (top panels) and $10^{10}\msun$ (bottom panels), respectively. The symbols in each panel are the measurements in Figure~\ref{fig:wp_MHI}, while the red and blue solid lines are the corresponding models with the minimum and maximum values of large-scale bias (see the text for details).
	} \label{fig:shampar}
\end{figure*}

At a fixed halo mass, halo clustering depends on the assembly history (known as the assembly bias), as originally found by \cite{Gao05} using the Millennium Simulation \citep{Springel05b} and confirmed by many follow-up studies \citep[see e.g.,][]{Zhu06,Wechsler06,Gao07,Jing07,Wang07,Dalal08,Li08,Lacerna11,Hearin16}. Halo parameters related to halo assembly history include halo formation time, halo concentration, and halo spin parameter. For halos at a fixed mass below the nonlinear mass for collapse (about $6.5\times 10^{12}\msun$ for our adopted cosmology), those with earlier formation time, higher concentration, or higher spin are more strongly clustered \citep[e.g.,][]{Wechsler06,Gao07,Jing07}. In fact, halo formation time is adopted in the aforementioned subhalo age distribution matching (SADM) model as a parameter to characterize the epoch when the cold gas supply of star formation is deprived and explain the color dependence of clustering for galaxies at fixed luminosity or stellar mass \citep{Hearin-Watson-13}. While halo formation time can be well considered as an additional parameter in this work, we also make attempts with halo concentration and spin. 

The formation time of a halo or subhalo is defined as the redshift $z_{1/2}$ at which the halo mass first reaches half of the peak value over the whole merger history. The concentration parameter $c$ is given by the ratio between the virial radius $R_{\rm vir}$ and the scale radius $R_{\rm s}$ of the halo. The halo spin parameter $\lambda$ is defined as \citep{Peebles69},
\begin{equation}
\lambda=\frac{J|E|^{1/2}}{GM_{\rm h}^{5/2}},
\end{equation}
where $J$ and $E$ are the magnitude of the angular momentum and the total energy of the halo. All the three parameters are available in the catalog of \texttt{ROCKSTAR} halos \citep{Behroozi13c} for the simulation we use.

We extend the SHAM model by including one of the three halo assembly parameters in addition to the peak circular velocity $V_{\rm peak}$. Given that the data do not require a sophisticated model yet, we develop a simple procedure to perform the modeling with the extended SHAM. For a given \hi-selected galaxy sample, we first preselect halos/subhalos according to the chosen halo assembly parameter. Then with such a selected sub-population of halos we perform the abundance matching with $V_{\rm peak}$ (assuming no scatter). The halos/subhalos are preselected by imposing an upper bound for the assembly parameter ($z_{1/2}$, $c$, or $\lambda$), as the samples are essentially in the halo mass regime that halos with a lower assembly parameter show weaker clustering. By selecting halos with a low assembly parameter, it is possible to cross the plateau in the halo bias-mass relation and thus to explain the low clustering amplitude in the low $\mhi$-threshold samples. 

In our simple model, the only free parameter is the upper bound for the chosen halo assembly parameter. We note that the change in halo bias by varying this upper bound comes from a combination of the assembly effect and $V_{\rm peak}$ effect on halo bias. For example, halos of a higher mass on average form later and cluster more strongly. If we lower the upper bound on $z_{1/2}$, we select low mass halos with lower bias (caused by the assembly effect), and at the same time, we also select relatively more halos of higher $V_{\rm peak}$ (higher bias) as they form later. Lowering the bias by the assembly effect by lowering the $z_{1/2}$ threshold can eventually be balanced by the inclusion of relatively more halos of higher $V_{\rm peak}$, leading to a lower limit for the clustering amplitude in the model. Similarly the model also has an upper limit for the clustering amplitude. Our model can be regarded as introducing a specific way of populating galaxies into halos as a function of $V_{\rm peak}$ and the chosen assembly parameter.
\begin{figure*}
	\centering
	\includegraphics[width=\textwidth]{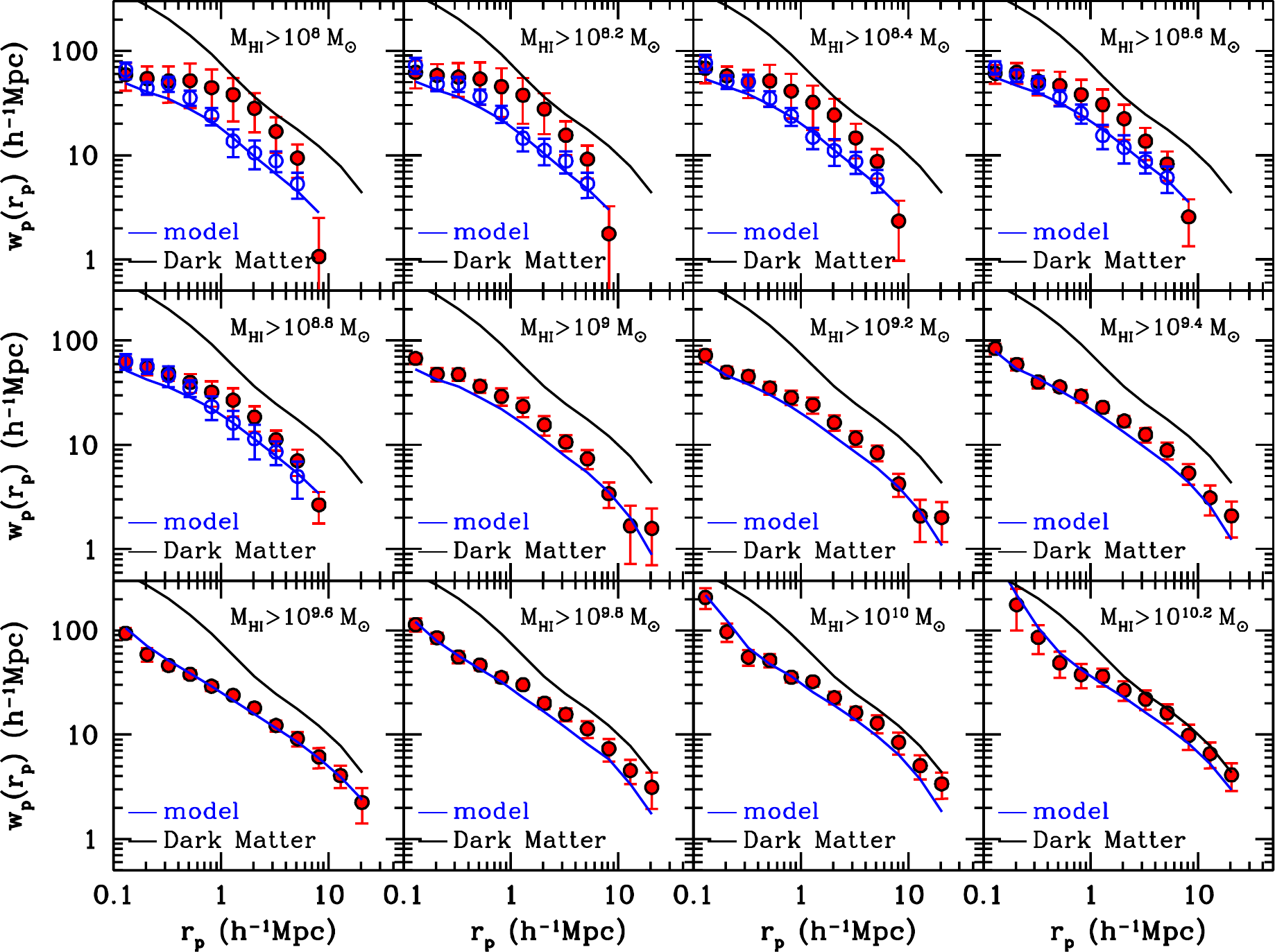}
	\caption{Best-fitting model predictions (blue solid lines) of $w_{\rm p}(r_{\rm p})$ for the different $\mhi$ threshold samples as in Figure~\ref{fig:wp_MHI}. We also show the measurements of $w_{\rm p}(r_{\rm p})$ from the SGC as the blue open circles in each panel of the samples with the \hi\ mass threshold lower than $10^9\msun$, where the effect of integral constraint has been corrected using the differences between the mock measurements in the SGC and the whole sample. .
	} \label{fig:wp_hod}
\end{figure*}

In Figure~\ref{fig:shampar} we show the $\wprp$ from the two extreme models (red and blue curves) with each of the three assembly parameters ($z_{1/2}$, $c$, or $\lambda$), predicting the minimum and maximum large-scale biases at a given $\mhi$ threshold ($\mhi>10^9\msun$ and $10^{10}\msun$ in the top and bottom panels).  The data points in each panel show the measurements (as in Figure~\ref{fig:wp_MHI}). Qualitatively, the clustering in the model is considerably weaker for smaller thresholds of $z_{1/2}$, $c$, or $\lambda$, and the effect is seen on all scales and for all the three halo assembly parameters. This is consistent with the assembly bias effect. Quantitatively, the model based on the halo formation time ($z_{1/2}$) can reasonably well fit the clustering measurements of both \hi-selected galaxy samples. The model with the concentration parameter reproduces the clustering at intermediate-to-small scales for both samples, but predicts too high clustering amplitudes at scales above a few Mpc for the low $\mhi$ sample. The model with the spin parameter generally fails to reproduce the data at all scales and \hi\ masses. Given the results from the above investigation, in what follows we limit our study to the model with halo formation time.

\subsection{The SHAM model with halo formation time thresholds}
\label{sec:sham}

For each of the 13 $\mhi$-threshold samples, we obtain the best-fitting threshold of halo formation time $z_{1/2}$ by minimizing the $\chi^2$,
\begin{equation}
\chi^2= (\mathbf{w_{\rm p}}-\mathbf{w_{\rm p}^*})^T \mathbf{C}^{-1}(\mathbf{w_{\rm p}}-\mathbf{w_{\rm p}^*}), \label{eq:chi2wp}
\end{equation}
where $\mathbf{C}$ is the full error covariance matrix of $w_{\rm p}(r_{\rm p})$. As we only have one free parameter, the degree of freedom (dof) is calculated as, ${\rm dof}=N_{\rm data}-1$, where $N_{\rm data}$ is the number of data points in the $\wprp$ measurement. 

Since the $\mhi$-threshold samples are not independent of each other by selection, the modeling cannot be done independently for each sample. We adopt a straightforward method that models the different $\mhi$-threshold samples coherently. We start with the most massive $\mhi$ threshold sample, applying our model to obtain a best-fitting $z_{1/2}$ threshold for the sample, and we assign an \hi\ mass to each halo or subhalo that has a formation time later than the $z_{1/2}$ threshold by abundance matching $\mhi$ to $V_{\rm peak}$. We then successively apply the same procedure for lower $\mhi$-threshold samples. For each sample, we keep the \hi\ mass fixed for halos that are already assigned an \hi\ mass in the previous samples, thus assigning an \hi\ mass only to halos that are additionally selected by the $z_{1/2}$ threshold of the current sample. After this process is done for all the samples, we essentially obtain a mock galaxy catalog in the simulation box which reproduces the clustering of all the \hi-selected galaxies with $\mhi>10^8\msun$. 

In order to appropriately take into account the effect of integral constraint, we have constructed 64 mock catalogs that have the same survey geometry as the ALFALFA sample. To this end, we divide the simulation box into 64 sub-volumes, place the virtual observer at the center of each sub-volume, and construct a mock catalog out of each sub-volume by applying our best-fit model. The line-of-sight velocity of each (sub)halo hosting a mock galaxy is used to account for the redshift-space distortion effect. We have also applied the $W_{50}$-dependent line flux limit $S_{\rm int}$ in ALFALFA by using the \hi\ mass and the associated line profile width $W_{50}$ for each mock galaxy. The average $w_{\rm p}(r_{\rm p})$ from the 64 mock catalogs is adopted as the model prediction for a given \hi\ mass threshold.

\begin{figure*}
	\centering
	\includegraphics[width=0.8\textwidth]{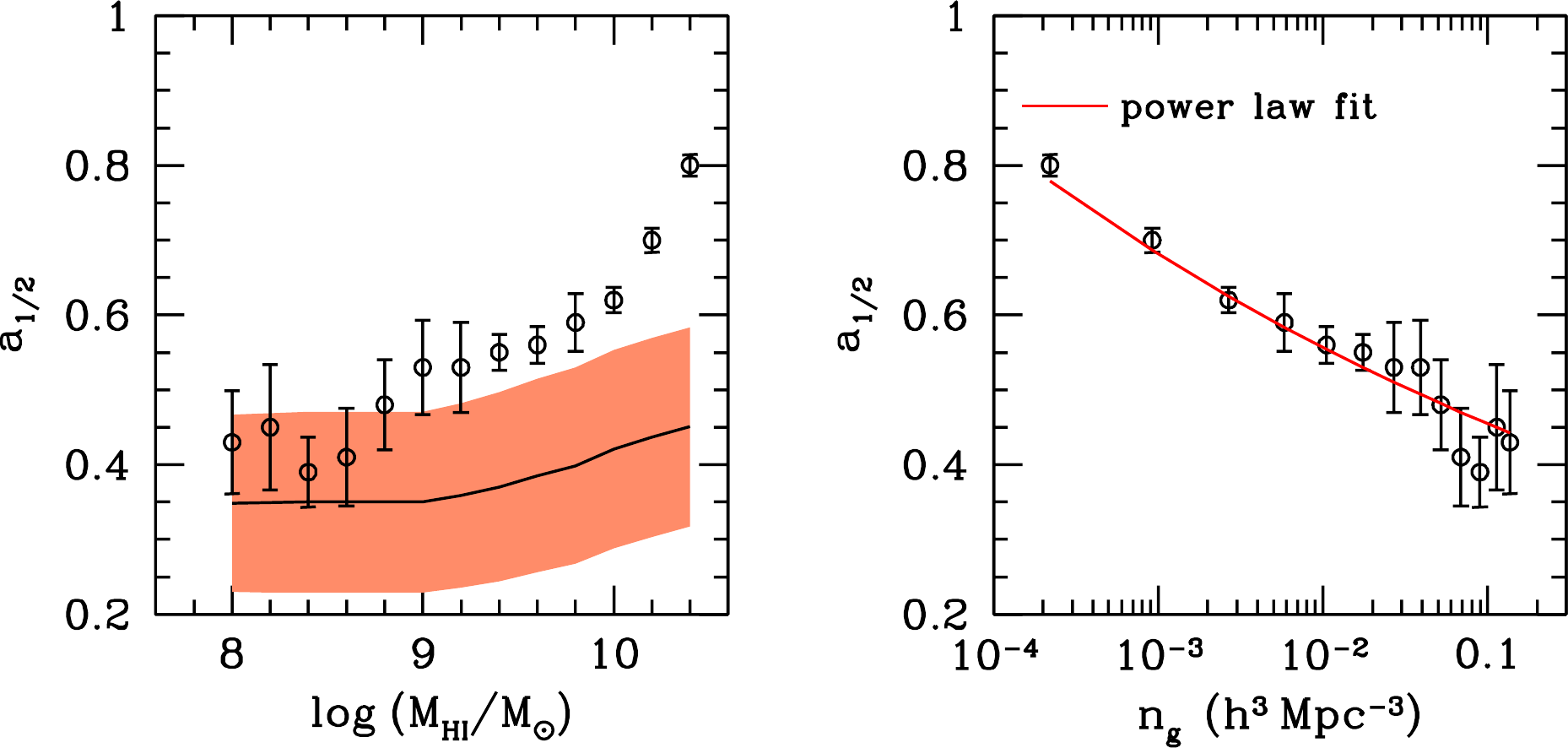}
	\caption{Left: halo formation time threshold $a_{1/2}$ as a function of the $\mhi$ thresholds of the galaxy samples. The circles show the best-fitting model predictions. The black solid line is the average halo formation time for the halos selected with the same $V_{\rm peak}$ cut as the best-fitting models but without applying the $a_{1/2}$ cut. The shaded area is the $1\sigma$ scatter around the mean. Right: halo formation time threshold $a_{1/2}$ as a function of galaxy number density. The red solid line is a power law fit of $a_{1/2}=0.682\,[n_{\rm g}/(10^{-3}\,h^3\,{\rm Mpc}^{-3})]^{-0.088}$.
	} \label{fig:ahalf}
\end{figure*}

Figure~\ref{fig:wp_hod} compares the observed $w_{\rm p}(r_{\rm p})$ (red symbols with error bars) with the model prediction (blue solid lines) for 12 $\mhi$-threshold samples. For comparison, we show the measurements of $w_{\rm p}(r_{\rm p})$ from the SGC alone, as the blue symbols with error bars, for which we have corrected the effect of the integral constraint according to the average correction estimated from the mock catalogs (see \S~\ref{sec:mocks}). The model reproduces the observed $\wprp$ for \hi-selected samples with thresholds above $\mhi\sim10^9\msun$. For lower masses, the data points are systematically higher than the model at intermediate scales, which can be largely attributed to the sample variance in the observational sample as discussed in the previous section. Thus, the discrepancy between the data and model at these low masses should not be overemphasized given the large errors in the observed $\wprp$. Because of the large errors the model parameters at the low masses are poorly constrained, as we will discuss below. 

It is interesting to see that, although the model parameters are constrained by the whole sample at a given $\mhi$ threshold, the predicted $\wprp$ appears to match very well with the data points from the SGC. In fact, we have tested that both the best-fit model parameters and the predicted $\wprp$ remain almost unchanged if we perform the fit with only the SGC data instead of the whole sample for the \hi\ mass thresholds below $10^9\msun$. As we use a global halo population in our model, it cannot offer the flexibility to account for the sample variance effect, and this contributes to the insensitivity of the results to the measurements. The sample variance effect should also be reflected in the covariance matrix, which can further contribute to the above insensitivity. That is, in contrast to the measurements, the modeling results are less affected by the sample variance effect. On the other hand, the agreement between the model and the SGC data reinforces the conjecture that the SGC is less affected by sample variance and provides more reasonable measurements of $\wprp$ at low \hi\ masses.

Table~\ref{tab:hod} lists the best-fitting model parameters for all the $\mhi$-threshold samples, including the \hi-mass threshold of the sample, the reduced $\chi^2$, the minimum value of $V_{\rm peak}$, the halo formation time threshold and the satellite fraction $f_{\rm sat}$. The halo formation time threshold is given in terms of the scale factor, $a_{1/2}\equiv1/(1+z_{1/2})$. The errors on the halo formation time is given by the variation of $z_{1/2}$ around the best-fitting value with $\Delta\chi^2=1$. The satellite fraction is around $10\%$, but slightly larger at the lowest masses.

\begin{table}
	\centering
	\caption{Best-fitting models for different $\mhi$-threshold samples} \label{tab:hod}
	\begin{tabular}{lrrrr}
		\hline
		Sample  & $\chi^2/\rm{dof}$ & $V_{\rm peak}$ & $a_{1/2}$ & $f_{\rm sat}$ (\%)  \\
		\hline			
		$\mhi>10^{8.0}M_{\odot}$ & 8.64/9 & 37.38 & $0.43\pm0.07$ & 11.45\\
		$\mhi>10^{8.2}M_{\odot}$ & 6.80/9 & 39.87 & $0.45\pm0.08$ & 11.92 \\
		$\mhi>10^{8.4}M_{\odot}$ & 4.53/9 & 41.56 & $0.39\pm0.04$ & 12.48 \\
		$\mhi>10^{8.6}M_{\odot}$ & 5.96/9 & 41.56 & $0.41\pm0.06$ & 12.11 \\
		$\mhi>10^{8.8}M_{\odot}$ & 7.36/9 & 41.56 & $0.48\pm0.06$ & 11.37 \\
		$\mhi>10^{9.0}M_{\odot}$ & 7.97/11 & 41.56 & $0.53\pm0.06$ & 11.02 \\
		$\mhi>10^{9.2}M_{\odot}$ & 7.31/11 & 50.44 & $0.53\pm0.06$ & 10.85 \\
		$\mhi>10^{9.4}M_{\odot}$ & 9.34/11 & 60.62 & $0.55\pm0.02$ & 10.63 \\
		$\mhi>10^{9.6}M_{\odot}$ & 13.06/11 & 73.87 & $0.56\pm0.02$ & 10.54 \\
		$\mhi>10^{9.8}M_{\odot}$ & 10.90/11 & 86.41 & $0.59\pm0.04$ & 10.23 \\
		$\mhi>10^{10.0}M_{\odot}$& 12.10/11 & 110.69 & $0.62\pm0.02$ & 9.13 \\
		$\mhi>10^{10.2}M_{\odot}$& 6.85/10 & 131.14 & $0.70\pm0.02$ & 8.23 \\
		$\mhi>10^{10.4}M_{\odot}$& 14.37/10 & 150.78 & $0.80\pm0.01$ & 10.29 \\		
		\hline
	\end{tabular}
	
	\medskip
	The best-fitting halo formation time $a_{1/2}$ (in terms of scale factor) for different $\mhi$-threshold samples. The $\chi^2/\rm{dof}$, minimum value of $V_{\rm peak}$ (in units of $\kms$) and satellite fraction $f_{\rm sat}$ for each model are also shown.
\end{table}

Figure~\ref{fig:ahalf} (left panel) displays the halo formation time threshold $a_{1/2}$ as a function of the $\mhi$ threshold. Generally, $a_{1/2}$ increases with increasing $\mhi$, ranging from $\sim0.4$ for $\mhi>10^8\msun$ up to $\sim0.8$ for $\mhi>10^{10.4}\msun$ which correspond to $z_{1/2}\sim1.5$ and $\sim0.25$, respectively. The halo mass dependence of $a_{1/2}$ is relatively weak at masses below $\sim10^{10}\msun$ and becomes strong at high masses. For comparison we show the average value of $a_{1/2}$ and the $1\sigma$ scatter for halos selected with the same $V_{\rm peak}$ cuts as in our model but without the additional selection by $a_{1/2}$. Overall, our model selects younger-than-average halos to host the \hi-selected galaxies, with larger differences in $a_{1/2}$ at higher \hi\ masses. In the right panel of the same figure we show the $a_{1/2}$ threshold as a function of the galaxy number density predicted by our model, which can be described by a power-law relation of $a_{1/2}=0.682\,[n_{\rm g}/(10^{-3}\,h^3\,{\rm Mpc}^{-3})]^{-0.088}$. We note that in this fitting formula, $a_{1/2}$ exceeds unity at $n_{\rm g}<1.3\times10^{-5}\,h^3\,{\rm Mpc}^{-3}$, thus the formula is not applicable for extremely low densities. 

\section{Model implications and comparisons with the literature} \label{sec:hihm}

With our bestfit model, we can derive various properties and  relationships for \hi-selected galaxies. We compare the predictions from our model with those from previous work based on different methodologies.

\subsection{\hi-to-halo mass relation}

\begin{figure}
	\centering
	\includegraphics[width=0.45\textwidth]{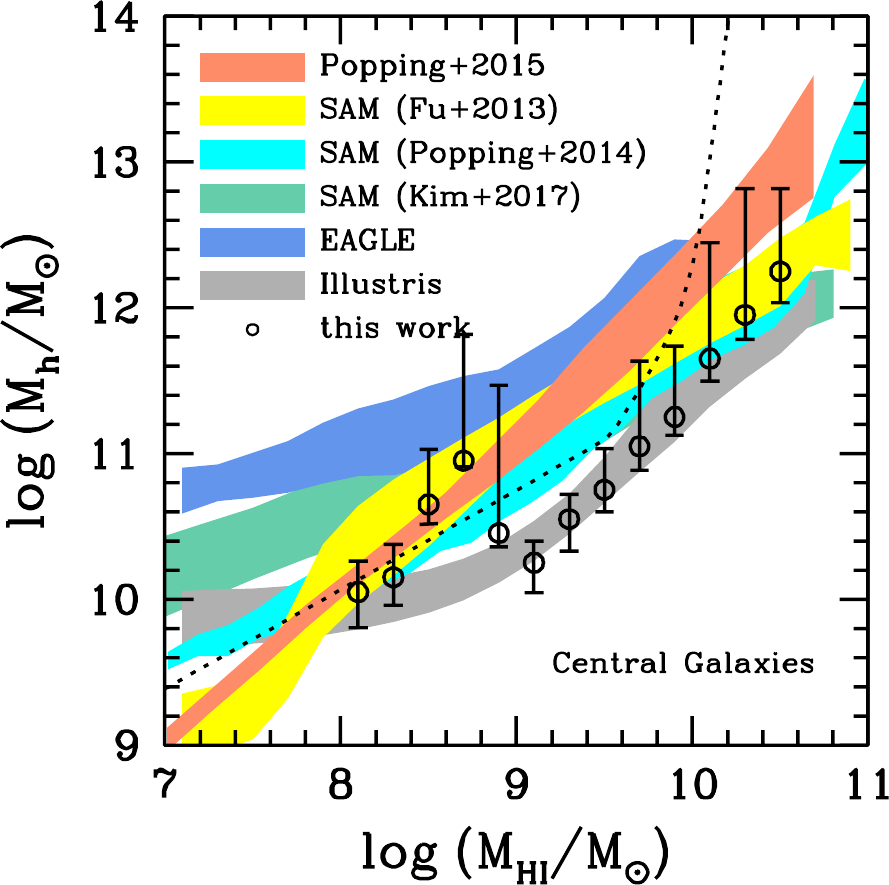}
	\caption{\hi-halo mass relation for central galaxies. We compare our best-fitting model (open circles with error bars) with six different models, including a halo-based semi-empirical model of \cite{Popping15}, three semi-analytical models (SAMs) of \cite{Fu13}, \cite{Popping14b}, and \cite{Kim17}, and two hydrodynamical simulation models from EAGLE \citep[][]{Crain17} and Illustris \citep[][]{Vogelsberger14b}. For convenience, we display the $1\sigma$ ranges of the \hi-halo mass relations from different models as color bands. The dotted line is the prediction from combining the average \hi-stellar and stellar-halo mass relations (see text). 
	} \label{fig:hihm}
\end{figure}

Our modeling results give the galaxy-halo relation for a series of $\mhi$-threshold samples, from which we can construct the galaxy-halo relation from galaxies in narrow $\mhi$ bins. Figure~\ref{fig:hihm} displays the $\mhi$ versus halo mass relation as inferred for the central galaxies in our best-fitting model (circles with error bars). We use the halo mass at which the distribution of $M_{\rm h}$ is peaked as the typical halo mass for each $\mhi$ bin. The error bars on $M_{\rm h}$ indicate the mass range of halos that host $68.3\%$ of the \hi\ galaxies in the $\mhi$ bin. The large upper error bar, as seen particularly at large $\mhi$, indicate both that the host halos of \hi-rich galaxies span a wide range of halo mass and that \hi-rich galaxies are preferentially found in relatively low-mass halos at a given $\mhi$. 

A noticeable result from our model is the non-monotonic relation between $M_{\rm h}$ and $\mhi$, in the sense that $M_{\rm h}$ increases with $\mhi$ but drops suddenly at $\mhi\sim10^{8.7}\msun$, before increasing again at $\mhi\sim10^{9.1}\msun$. The overall relationship can be accurately described by a triple power law function as follows,
\begin{equation}
     \log M_{\rm h} = \left\{
     \begin{array}{lr}
       1.38 \log \mhi-1.15, \,\, &  \log \mhi < 8.7, \\
       25.40-1.67 \log \mhi, \,\, & 8.7\leq \log \mhi \leq 9.1, \\ 
       1.38 \log \mhi-2.32, \,\, &  \log \mhi > 9.1.
     \end{array}
   \right.
\end{equation}

The \hi-halo mass relation may be empirically estimated from combining the \hi-stellar mass relation \citep{Huang12} with the stellar-halo mass relation \citep{Moster10}, both available from previous studies. The result is shown as the dotted line in Figure~\ref{fig:hihm}, which is a monotonically increasing function with relatively flat (steep) slope below (above) $\mhi\sim10^{9.5}\msun$. This relation agrees roughly but not exactly with our model for $\mhi<10^{9.5}\msun$ in terms of both amplitude and slope. At higher \hi\ masses, this empirical relation predicts much higher halo masses than  our model. This may be attributed to the scatter in the \hi-to-stellar mass and stellar-to-halo mass relations which are not considered here. The steep relation at the high-mass end is expected to flatten if the scatter is included, thus becoming more consistent with our model prediction.  

The \hi-to-halo mass relation has been obtained in previous studies in different ways. For comparison, Figure~\ref{fig:hihm} shows the \hi-halo mass relations predicted by six different models. These  include the halo-based semi-empirical model by \cite{Popping15}, three different semi-analytical models (SAMs) of galaxy formation by \cite{Fu13}, \cite{Popping14b} and \cite{Kim17}, and two hydrodynamical simulations by \cite{Crain17} (the EAGLE project) and \cite{Vogelsberger14b} (the Illustris project). For clarity, we only show the $1\sigma$ range and omit the average relation for each model. We have corrected the effect of different cosmological parameters adopted in the different models, scaling the halo masses by the ratio of $\Omega_{\rm m}$ and the \hi\ masses by the ratio of $\Omega_{\rm b}$, both relative to the value adopted in this work. 

The three SAMs are built upon the halo merger trees from high-resolution $N$-body simulations and include the partitioning of cold gas into atomic and molecular phases by applying empirical prescriptions. It is thus not surprising to see the \hi-to-halo mass relations from these models being quite similar to each other at \hi\ masses above $\sim10^8\msun$. At the low-mass end, the SAMs of \cite{Fu13} and \cite{Popping14b} agree well, but the SAM of \cite{Kim17} predicts significantly higher halo masses, which may be attributed to the photoionization feedback additionally implemented in order to bring the model to better match the low-mass end of the observed \hi\ mass function (HIMF). The empirical relation from \cite{Popping15} was obtained for galaxies at $z=0$ in two successive steps: first assigning a star formation rate (SFR) to each galaxy using the fitting functions from \cite{Behroozi13}, and then estimating both \hi\ and H$_2$ masses by the combination of an empirical molecular gas-based SFR relation and a pressure-based molecular gas fraction relation. This result agrees broadly with the SAMs at \hi\ masses below $\sim10^{10}\msun$. At higher masses, the halo mass predicted by \cite{Popping15} increases more rapidly with increasing $\mhi$. The authors considered the scatter in the SFR-stellar mass relation but ignored the scatter in the gas-related relations, which might be part of the reason for the steep \hi-to-halo mass relation at high masses. 

The two hydrodynamic simulations present similar slopes, but significantly different amplitudes in the \hi-to-halo mass relation. For $\mhi<10^9\msun$ and at given $\mhi$, the halo mass of \hi-selected galaxies in the EAGLE simulation is an order of magnitude higher than the halo mass in the Illustris simulation. In other words, at a fixed halo mass, the EAGLE simulation predicts too low \hi\ mass in galaxies. At $\mhi>10^9\msun$ the difference decreases with increasing $\mhi$,  but still it is as large as $\sim0.5$ dex even at $\mhi\sim10^{10}\msun$. The SAMs and the empirical model of \cite{Popping15} fall in between the EAGLE and Illustris simulations. More work is needed if one were to fully understand the reason behind the large discrepancy between the two simulations, but the relatively poor resolution of the EAGLE simulation could be part of the reason. Empirical models rather than physics models related to cold gas have had to be applied in the EAGLE simulation given the limited resolution, resulting in an HIMF significantly lower than the observed one, as shown in \cite{Crain17} (see their Fig.~8 and also Fig.~\ref{fig:himf} below).

Our model agrees quite well with the Illustris simulation for $\mhi>10^9\msun$ where the clustering measurements are robust to the sample variance (see discussion in \S~\ref{sec:measurements}). At $\mhi<10^{8.7}\msun$, the Illustris simulation predicts a nearly constant halo mass, independent of $\mhi$. In contrast, the halo mass from our model still strongly depends on $\mhi$ at these low \hi\ masses, a behavior which agrees with both the empirical model of \cite{Popping15} and the SAMs of \cite{Fu13} and \cite{Popping14b}. However, we would like to point out that, the agreement with these models at the low mass end should not be overemphasized, again given the large uncertainties in the clustering measurements as extensively discussed above. Larger and deeper \hi\ surveys are needed to reliably determine the low-mass end of the \hi-to-halo mass relation. 

\subsection{Halo occupation distributions}

\begin{figure}
	\centering
	\includegraphics[width=0.45\textwidth]{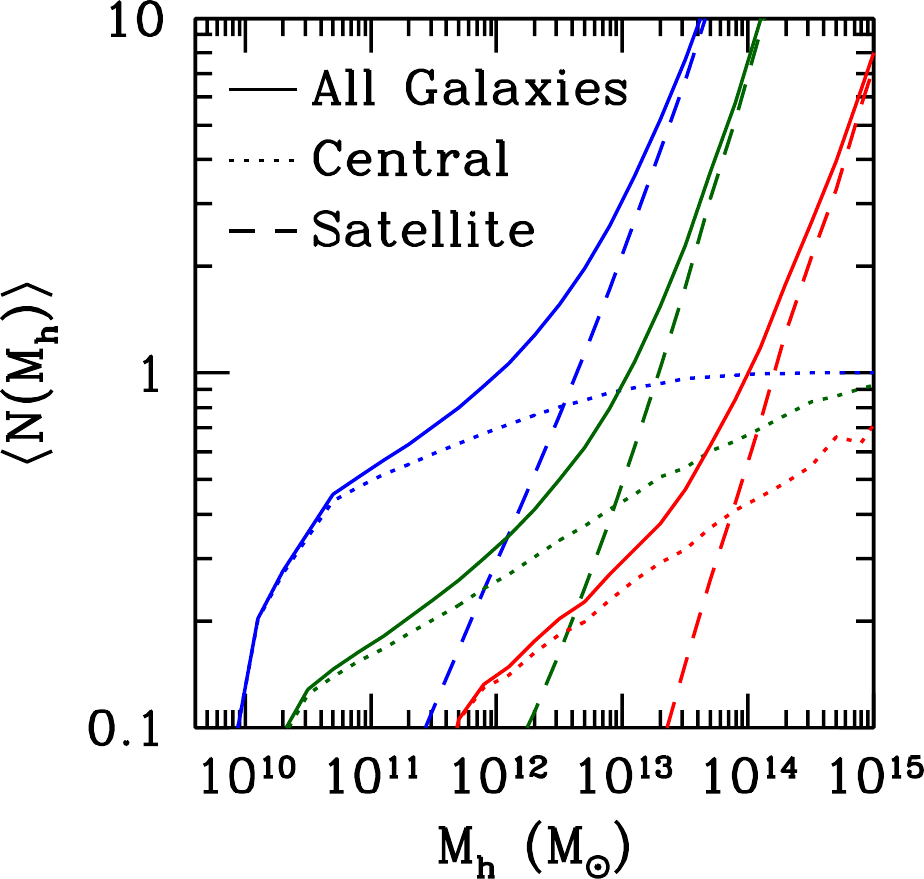}
	\caption{Mean occupation functions from the best-fitting models for the three typical $\mhi$-threshold samples with $\mhi>10^8\msun$ (blue lines), $10^9\msun$ (green lines), and $10^{10}\msun$ (red lines). The total mean halo occupation function (solid lines) is decomposed into contributions from central galaxies (dotted lines) and satellite galaxies (dashed lines).
	} \label{fig:hod}
\end{figure}

\begin{figure*}
	\centering
	\includegraphics[width=0.8\textwidth]{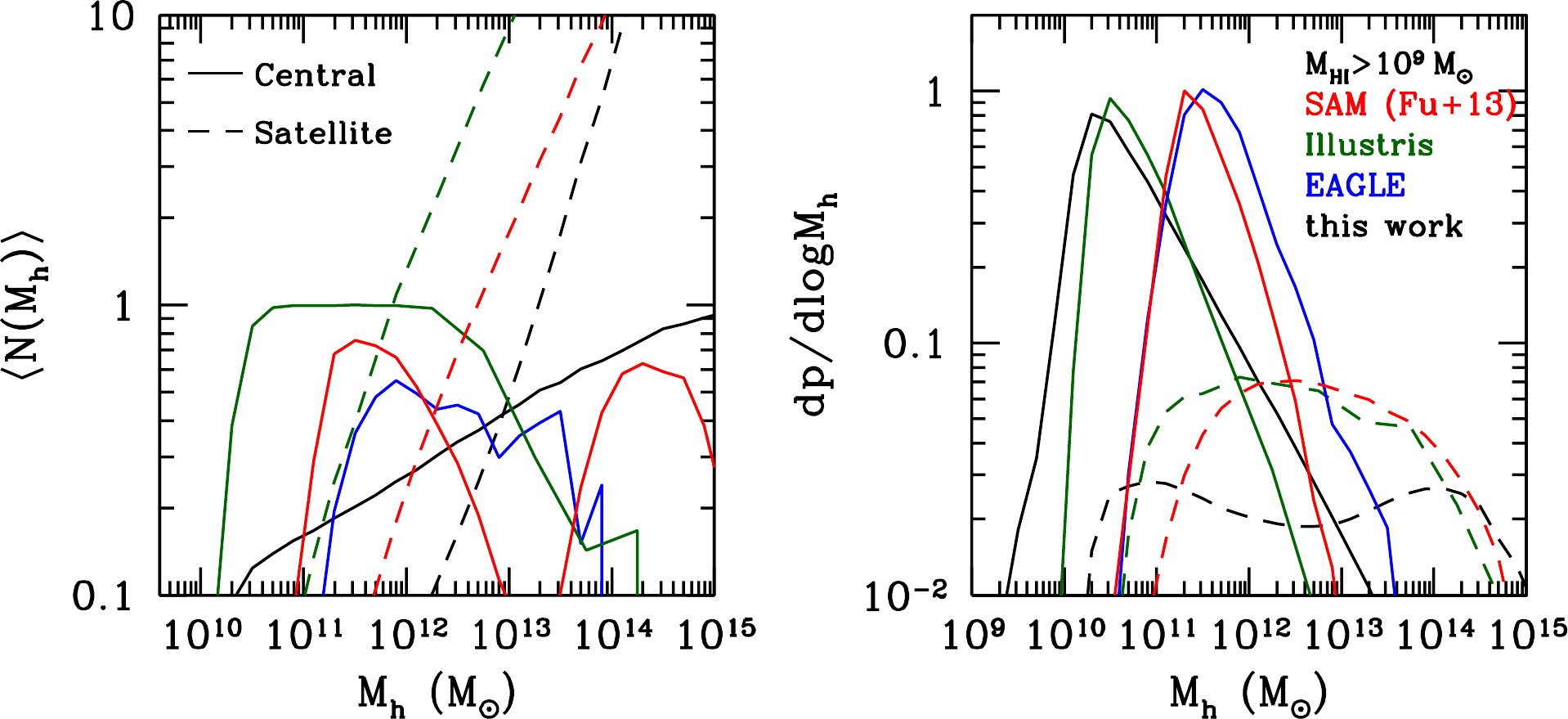}
	\caption{Mean halo occupation number $\langle N(M_{\rm h})\rangle$ (left panel) and the differential probability distribution of halo mass $dp/d\log M_{\rm h}$ (right panel) for the sample of $\mhi>10^9\msun$. The predictions from our best-fitting model are shown as the black lines, while the results from the SAM of \cite{Fu13}, results from EAGLE and Illustris simulations are shown as the red, blue, and green lines, respectively. The models for the central and satellite galaxies are displayed in solid and dashed curves, respectively. } \label{fig:hod_hihm}
\end{figure*}

Figure~\ref{fig:hod} displays the mean occupation distribution functions inferred from our best-fitting models for three typical $\mhi$-threshold samples with $\mhi>10^8\msun$ (blue lines), $10^9\msun$ (green lines), and $10^{10}\msun$ (red lines). The total mean halo occupation function, $\langle N(M_{\rm h})\rangle$ (solid lines), is decomposed into contributions from central galaxies, $\langle N_{\rm cen}\rangle$ (dotted lines), and satellite galaxies, $\langle N_{\rm sat}\rangle$ (dashed lines). Because we have only selected halos younger than the formation time threshold $a_{1/2}$, it is not surprising that the central occupation numbers for the samples of $\mhi>10^9$ and $10^{10}\msun$ are smaller than unity even for very massive halos with $M_{\rm h}\sim10^{15}\msun$. 

In traditional HOD models, the characteristic host halo mass for central galaxies with a threshold luminosity or stellar mass is usually given by $M_{\rm min}$, the mass at which the mean occupation number of central galaxies per halo is 0.5  \citep[see Eq.~\ref{eq:Ncen};][]{Zehavi05,Zehavi11,Zheng09,White11,Guo14,Guo15c}. In our case, however, the central galaxy occupation does not show a rapid cutoff at the low mass end, particularly for the samples with $\mhi>10^9\msun$ or higher. For instance, for the sample of $\mhi>10^{10}\msun$, $M_{\rm min}\sim3\times10^{14}\msun$, but most of the central galaxies in this sample live in halos of about $10^{11.5}\msun$ (see Fig.~\ref{fig:hihm}). This results from the fact that, by construction, our model tends to put \hi\ galaxies into young, low mass halos to reduce the large-scale bias. It should be noted that the mean occupation function implied by our model is of limited use for the general HOD modeling, as halos are pre-selected according to their assembly history, breaking the HOD assumption that the galaxy content depends only on halo mass.

Figure~\ref{fig:hod_hihm} shows the average halo occupation number $\langle N(M_{\rm h})\rangle$ (left panel) and the differential probability distribution of halo mass $dp/d\log M_{\rm h}$ (right panel) for the $\mhi>10^9\msun$ sample. The probability distribution of halo mass is derived from the product of $\langle N(M_{\rm h})\rangle$ and the halo mass function. We show the results only for the sample with $\mhi>10^9\msun$ and compare these with the predictions of the same quantities from three models: the SAM of \cite{Fu13}, the EAGLE simulation by \cite{Crain17} and the Illustris simulation by \cite{Vogelsberger14b}. Results are shown separately for central and satellite galaxies. We consider only the three models because data for other models are not available to us. We note that the EAGLE simulation data from \cite{Crain17} only include the central galaxy population.

For central galaxies, the SAM of \cite{Fu13} predicts a bimodal distribution of $\langle N_{\rm cen}(M_{\rm h})\rangle$, with two well-separated populations peaked at $M_{\rm h}\sim10^{11.5}\msun$ and $10^{14.5}\msun$, respectively. This bimodal distribution is not seen in the hydrodynamic simulations in which the \hi-selected central galaxies are limited to halos with intermediate-to-low masses. Neither is the bimodal distribution seen in our model, but our model appears to be in better agreement with the SAM of \cite{Fu13}, in the sense that both models predict a considerable fraction of the most massive halos to be able to host \hi-rich galaxies. For satellite galaxies, all the models predict a power-law mean occupation function with halos of higher masses hosting larger numbers of \hi-selected galaxies, but on average our model requires the host halos of satellites to be more massive than those in the other models. In our model, the existence of \hi-rich satellites in massive halos is possible as long as the halos are formed at a substantially late time. This population must be very rare in observational samples given the extremely low abundance of the most massive halos. In fact, \hi\ emission has been detected in satellite galaxies of massive halos with $M_{\rm h}\sim10^{14}-10^{15}\msun$ \citep[e.g.][]{Catinella13}.  

\begin{figure*}
	\centering
	\includegraphics[width=0.9\textwidth]{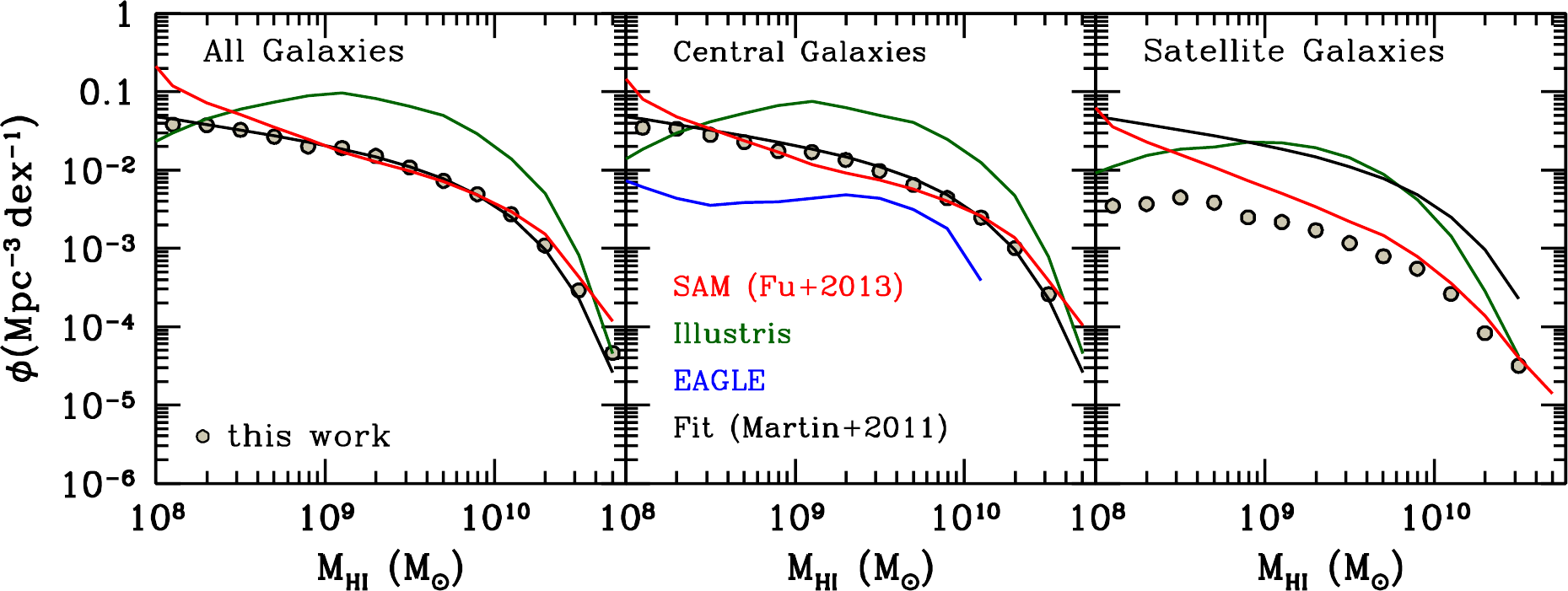}
	\caption{\hi\ mass functions of all (left panel), central (middle panel), and satellite galaxies (right panel) from several models. The results from the SAM of \cite{Fu13}, Illustris, and EAGLE are displayed as the red, green, and blue lines, respectively. Our model predictions are shown as the filled circles. We also show for comparison the Schechter function fit from \cite{Martin10} for all galaxies in the ALFALFA sample as the black line in each panel.} \label{fig:himf}
\end{figure*}

The actual shape of the occupation function at high-mass does not have a significant effect on the probability distribution of host halo mass, because the halo mass function decreases rapidly with $M_{\rm h}$ at the massive end. This can be clearly seen from the right panel of Figure~\ref{fig:hod_hihm}, where the probability distribution function of host halo mass for central galaxies is concentrated to a narrow range of halo mass, ranging from a few $\times10^{9}\msun$ to $\sim10^{13}\msun$, with a peak at $\sim2\times10^{10}\msun$. This result is in good agreement with the Illustris simulation, while the distribution of halo mass for central galaxies in both the EAGLE simulation and the SAM of \cite{Fu13} covers a similarly narrow halo mass range but is peaked at $2-3\times10^{11}\msun$, an order of magnitude higher than predicted by our model. For satellite galaxies, all the models including ours span a broad range of halo mass from $M_{\rm h}\sim10^{11}\msun$ up to $\sim10^{15}\msun$, although our model shows a slightly different shape in the overall distribution.

\subsection{\hi\ mass function}

In Figure~\ref{fig:himf} we show the HIMF of galaxies as predicted by our model for the full galaxy population (left panel), and the populations of central (middle panel) and satellite (right panel) galaxies. For comparison, the Schechter function fit of the observational HIMF from ALFALFA \citep{Martin10} is plotted as a black solid line and repeated in every panel. By construction the HIMF from our model perfectly matches the observed HIMF for the whole \hi\ galaxy population. Furthermore, our model predicts that the overall HIMF is predominantly contributed by the central galaxy population. The central and satellite populations contribute about 90\% and 10\% of the total abundance at a given $\mhi$, and this result is essentially independent of the \hi\ mass, broadly consistent with the satellite fractions listed in Table~\ref{tab:hod} which are constant at $\sim10\%$.

Figure~\ref{fig:himf} also shows the HIMF prediction of the Illustris and EAGLE simulations, as well as the SAM of \cite{Fu13}. Despite the agreement on the \hi-halo mass relation between our model and the Illustris simulation for $\mhi>10^9\msun$ (see Fig.~\ref{fig:hihm}), the overall abundance of the \hi-selected galaxies from Illustris is significantly higher than the other models, with largest discrepancies at intermediate $\mhi$, and this is true for both centrals and satellites. The EAGLE simulation predicts an HIMF with a much lower amplitude than others, though the prediction is available only for central galaxies. The SAM of \cite{Fu13} appears to broadly agree with our model, particularly for the central galaxies and the full population. For satellite galaxies, the SAM predicts a steeper slope at the low-mass end, where the HIMF from our model is quite flat. It is not surprising that the SAM can successfully reproduce the overall HIMF as the model parameters were tuned to do so. It is encouraging that the SAM model and our model show a broad agreement in both the central and satellite components of the HIMF.

\section{Conclusions} \label{sec:conclusion}

In this paper, we have investigated the dependence of clustering on the \hi\ content of galaxies using the 70\% complete sample of the ALFALFA survey. We select galaxy samples by different \hi\ mass thresholds, ranging from $\mhi>10^8\msun$ to $\mhi>10^{10.4}\msun$, and for each sample we have estimated three clustering statistics: the projected 2PCF $\wprp$, the projected cross-correlation function with respect to a reference sample, and the redshift-space monopole moment $\xi_0(s)$. We construct a halo-based statistical model in which the \hi\ content of a galaxy depends on both the host dark matter halo mass and the halo formation time, and we use the $\wprp$ measurements to constrain the model parameters. We discuss the \hi-halo mass relation and \hi\ mass functions for central and satellite galaxies as inferred from our best-fitting models, and compare these results with predictions of the same statistics from other models in the literature including the traditional HOD models, semi-analytic models of galaxy formation, and cosmological hydro-dynamical simulations.

Our main conclusions can be summarized as follows.

\begin{itemize}
\item In contrast to previous studies that found no significant dependence of clustering on \hi\ mass, we find that the projected 2PCFs depend strongly on the \hi\ mass, in the sense that galaxies of higher \hi\ masses are more strongly clustered on scales above a few Mpc than the lower \hi\ mass galaxies. This finding is robust, as we also infer consistent galaxy bias factors from the redshift-space monopole moments and the cross-correlations between the ALFALFA and SDSS galaxy samples.  

\item The bias factors of the low \hi-mass samples are systematically lower than the minimum bias of dark matter halos selected by mass or $V_{\rm peak}$ thresholds. This implies that the relation between \hi-rich galaxies and halos depends not only on halo mass or $V_{\rm peak}$, as commonly assumed in traditional halo models such as the HOD model or the simplest SHAM model.

\item The clustering measurements of the \hi-selected samples can be reasonably explained by an extended SHAM model, which includes a parameter related to the halo assembly bias effect in addition to $V_{\rm peak}$. In our model, this parameter is chosen to be the halo formation time.
\end{itemize}

\section{Discussions} \label{sec:discussion}
Thanks to the most up-to-date ALFALFA $\alpha.70$ sample, we are able to perform an extensive investigation about the \hi-mass dependence of galaxy clustering. We make full use of the flux-limited sample through weighing galaxy pairs by $1/V_{\rm max}$ to improve the clustering measurements and to achieve effectively volume-limited measurements. We have tested the robustness of our clustering measurements by checking the finite volume effect, analyzing the NGC and SGC subsamples separately, measuring the clustering relative to a large reference sample, and carrying out comparisons with mock catalogs. We find that the clustering measurements at \hi\ masses below $10^9\msun$ are seriously biased by the existence of the super-clusters in the NGC, while the measurements based on the SGC data alone appear to be more consistent with the low bias factors at the low-mass end as obtained from the redshift-space monopole moments, which are less affected by the sample variance when compared to the projected 2PCFs. However, larger samples covering even larger volumes are still needed in order to have unbiased clustering measurements at these low \hi\ masses.

At $\mhi>10^9\msun$, where the different clustering statistics provide consistent results, we find the clustering on scales larger than a few Mpc to significantly depend on \hi\ mass, with stronger clustering at higher masses. Previous studies have controversial conclusions about this dependence \citep{Basilakos07,Meyer07,Papastergis13}, which can at least partially (if not purely) be attributed to the smaller sample sizes and the non-volume-limited clustering measurements. In fact, the \hi\ mass dependence of clustering is naturally expected given the positive correlation of $\mhi$ with stellar mass (e.g., Fig.2 of \citealt{Huang12}) and the known stellar mass dependence of clustering \citep[e.g.][]{Li06}. The stellar mass dependence has been studied in depth in the past decade for both low-$z$ galaxies and those at higher redshifts, thanks to the much larger samples from optical spectroscopic or photometric redshift surveys. These studies have provided stringent constraints on the link between galaxies of different stellar masses and their host dark matter halos. In principle, the \hi-mass dependence of clustering should also provide interesting constraints on the galaxy-halo relationship. Our halo-based model has shown that this is indeed the case.

The model we have proposed in this work is an extended version of the simple SHAM model. In our model the \hi\ content of a galaxy depends not only on the mass (or more accurately $V_{\rm peak}$) of its host halo as in the SHAM model, but also on the halo assembly history. It has been known that the clustering of halos depends not only on halo mass, but also on halo assembly history \citep{Gao05,Wechsler06,Gao07,Wang07,Li08}. If the galaxy formation and evolution tightly track the halo assembly, the above halo assembly bias effect would translate to an effect on galaxy clustering (a.k.a., galaxy assembly bias). We have adopted three parameters to quantify halo assembly history: halo formation time, halo spin, and halo concentration, and found the halo formation time to be the best parameter that can reproduce the clustering of \hi-rich galaxies at all scales. One interesting prediction of our model is that \hi-rich satellite galaxies could reside in massive halos as long as the halos are formed at substantially late times. This makes sense if the satellite galaxies lose their cold gas in a smooth manner. That is, the ram-pressure stripping of gas happens slowly, not immediately at accretion as assumed in most of the current semi-analytic models. Such smooth gas stripping has been observed in satellite galaxies in low-redshift galaxy clusters \citep[e.g.,][]{Zhang13}, and has recently been investigated in theoretical models \citep[e.g.,][]{Luo16}.

Recent studies of galaxy clustering have attempted to include the effect of halo assembly bias into the SHAM model. By tying galaxy color to halo age, such models can qualitatively reproduce the co-dependence of clustering on stellar mass and color \citep{Hearin-Watson-13,Zentner14,Zentner16}\footnote{We note that a recent study by \cite{Zu16} suggests that, at stellar masses above $10^{10}\,h^{-2}\,M_\odot$, models that tie galaxy color to halo age are disfavored by the observed strong halo mass bimodality between the red and blue central galaxies from weak lensing \citep{Mandelbaum16}.}. Apparently, our model is along the same line as these studies, and the results are not surprising given the close correlation between cold gas fractions of galaxies and their colors.

Observational studies attempting to directly detect the galaxy assembly bias in galaxy clustering have produced controversial results, however. On one hand, \cite{Miyatake16} and \cite{More16} analyzed two samples of galaxy clusters which have similar halo masses of $1.9\times10^{14}h^{-1}\,\msun$ as estimated from weak lensing signals, and found their halo bias values to differ by a factor of 1.5. This was regarded as observational evidence in support of the halo assembly bias effect. On the other hand, the assembly bias signal observed in the redMaPPer clusters is shown to be strongly contaminated by the projection effect of the cluster membership identification \citep[][]{Zu16b}. In addition, \cite{Lin16} compared the clustering of early- and late-forming galaxies with a mean halo mass of $\sim9\times10^{11}h^{-1}\msun$, and found no convincing evidence for galaxy assembly bias. In cosmological simulations, halo assembly bias is most pronounced for low-mass halos with masses below $\sim10^{13}\msun$, as originally found by \cite{Gao05}. Our model predicts that the \hi-rich galaxies detected in the ALFALFA are mostly hosted by halos less massive than $9\times10^{11}h^{-1}\,\msun$. Our results imply that the \hi\ content of galaxies may be more tightly correlated with the halo assembly history, compared to quantities derived from optical observations. Next-generation \hi\ surveys might be able to provide better constraints on the correlation between \hi\ gas and halo assembly, i.e., galaxy assembly bias on top of halo assembly bias.

We find that at $\mhi>10^{9}\msun$ the \hi-to-halo mass relation predicted by our model agrees well with the relation from the Illustris simulation, while the EAGLE simulation and current semi-analytic models predict higher halo masses at a given \hi\ mass. However, the Illustris simulation predicts an HIMF which is substantially too high when compared to the observed HIMF from ALFALFA. By construction our model accurately reproduces the observed HIMF for the full sample, and additionally predicts the HIMFs for central and satellite galaxies separately. It is interesting to note that the predicted central galaxy HIMF by our model agrees well with that from the SAM of \cite{Fu13}. For the satellite population, the same SAM agrees with our model at the high-mass end of the HIMF, but predicts too many galaxies at $\mhi<10^{10}\msun$. We did not examine the clustering properties predicted by these models, but in principle our measurements of the \hi-dependent clustering should be helpful for testing and constraining the models. In a recent paper, \cite{Zoldan17} have analyzed the \hi\ content of galaxies for six different SAMs, finding most of them to agree with the observational paper by \cite{Papastergis13} in the sense that both the models and the data reveal no significant mass dependence of the clustering at $\mhi>10^{9.5}\msun$. However, the results in \cite{Papastergis13} are based on a substantially smaller sample and on non-volume-limited clustering measurements. Given the \hi-mass dependence of the clustering we find here with larger samples and effectively volume-limited measurements, the comparison between the SAM predictions and the observational results needs to be revisited in order to properly test the models.  

With much larger samples of \hi-selected galaxies in the foreseeable future, we will be able to further verify our model predictions and study the relation between the cold gas and the halo environment in more depth. First, one may make use of optically-selected galaxy samples instead of \hi\ samples for the clustering analysis, estimating a {\it pseudo} \hi\ mass for each galaxy based on the tight scaling relations between the \hi\ gas content and the other properties such as color and surface brightness \citep[e.g.,][]{Zhang09, Li12, Teimoorinig17}. Current photometric \hi\ estimators, which are calibrated with the existing \hi\ surveys, such as ALFALFA and GASS, can provide unbiased \hi-to-stellar mass ratios for galaxies spanning a wide range of gas mass fraction, with scatter of only $\sim0.2-0.3$dex. By applying their estimator to a large sample of SDSS galaxies, \cite{Li12} examined the dependence of galaxy clustering on the \hi-to-stellar mass ratio and found that at a given stellar mass, galaxies with higher gas fractions have weaker clustering amplitudes. We expect that the poorly estimated clustering at the low \hi-mass end with the limited sample size to be well improved, thus providing additional and important constraints on our model, particularly at the low-mass end. The larger sample would also allow us to include not only the \hi\ mass, but also the \hi-to-stellar mass ratio in clustering measurements and modeling, a property supposed to be more closely linked to the halo assembly history. Second, we note that our simplified halo model is restrictive by construction as the accuracy of the clustering measurements from the small volume of the $\alpha$.70 sample does not allow for strong constraints on more sophisticated and flexible models. The traditional SHAM model usually includes the scatter in the stellar-to-halo mass as a free parameter \citep[see e.g.,][]{Behroozi10,Moster10,Leauthaud12,Nuza13,Reddick13}. In our model, we test the effect of the scatter but do not include it in the adopted model, although part of the scatter between the \hi\ and halo mass is taken into account through the scatter between $V_{\rm peak}$ and halo mass. In addition, in our model we assume a constant halo formation time threshold for each $\mhi$-threshold sample. With higher precision clustering measurements from future larger data, we expect to be able to better constrain the mapping of the galaxy \hi\ content onto the multi-dimensional halo parameter space (e.g., halo mass and formation time). 

\section*{Acknowledgements}

This work is supported by the National Key Basic Research Program of China (No. 2015CB857003, 2015CB857004). H.G. acknowledges the support of the 100 Talents Program of the Chinese Academy of Sciences and NSFC-11543003. C.L. acknowledges the financial support of NSFC (No. 11173045, 11233005, 11325314, 11320101002) and the Strategic Priority Research Program ``The Emergence of Cosmological Structures'' of CAS (Grant No. XDB09000000). H.J.M. acknowledges the support from NSF AST- 1517528 and NSFC-11673015. We thank the anonymous referee for the helpful comments that improve the presentation of this paper. We thank Jian Fu, Gerg\"o Popping, Robert A. Crain, Han-seek Kim, Emmanouil Papastergis and Michael G. Jones for kindly providing the data used in this paper.

We gratefully acknowledge the use of the High Performance Computing Resource in the Core Facility for Advanced Research Computing at Shanghai Astronomical Observatory. We acknowledge the Gauss Centre for Supercomputing e.V. (www.gauss-centre.eu) and the Partnership for Advanced Supercomputing in Europe (PRACE, www.prace-ri.eu) for funding the MultiDark simulation project by providing computing time on the GCS Supercomputer SuperMUC at Leibniz Supercomputing Centre (LRZ, www.lrz.de).


\begin{thebibliography}{144}
	
	\bibitem[{Abazajian} et al.(2009)]{Abazajian09}
	{Abazajian}, K.~N., {Adelman-McCarthy}, J.~K., {Ag{\"u}eros}, M.~A., et al.
	2009, \apjs, 182, 543
	
	\bibitem[{Basilakos} et al.(2007)]{Basilakos07}
	{Basilakos}, S., {Plionis}, M., {Kova{\v c}}, K., \& {Voglis}, N. 2007, \mnras,
	378, 301
	
	\bibitem[{Behroozi} et al.(2010)]{Behroozi10}
	{Behroozi}, P.~S., {Conroy}, C., \& {Wechsler}, R.~H. 2010, \apj, 717, 379
	
	\bibitem[{Behroozi} et al.(2013a)]{Behroozi13}
	{Behroozi}, P.~S., {Wechsler}, R.~H., \& {Conroy}, C. 2013a, \apj, 770, 57
	
	\bibitem[{Behroozi} et al.(2013b)]{Behroozi13c}
	{Behroozi}, P.~S., {Wechsler}, R.~H., {Wu}, H.-Y., et al. 2013b, \apj, 763, 18
	
	\bibitem[{Berlind} \& {Weinberg}(2002)]{Berlind02}
	{Berlind}, A.~A., \& {Weinberg}, D.~H. 2002, \apj, 575, 587
	
	\bibitem[{Blanton} \& {Roweis}(2007)]{Blanton07b}
	{Blanton}, M.~R., \& {Roweis}, S. 2007, \aj, 133, 734
	
	\bibitem[{Blanton} et al.(2005)]{Blanton05c}
	{Blanton}, M.~R., {Schlegel}, D.~J., {Strauss}, M.~A., et al. 2005, \aj, 129,
	2562
	
	\bibitem[{Blitz} \& {Rosolowsky}(2006)]{Blitz06}
	{Blitz}, L., \& {Rosolowsky}, E. 2006, \apj, 650, 933
	
	\bibitem[{Bothwell} et al.(2009)]{Bothwell09}
	{Bothwell}, M.~S., {Kennicutt}, R.~C., \& {Lee}, J.~C. 2009, \mnras, 400, 154
	
	\bibitem[{Bravo-Alfaro} et al.(2000)]{Bravo-Alfaro00}
	{Bravo-Alfaro}, H., {Cayatte}, V., {van Gorkom}, J.~H., \& {Balkowski}, C.
	2000, \aj, 119, 580
	
	\bibitem[{Catinella} et al.(2013)]{Catinella13}
	{Catinella}, B., {Schiminovich}, D., {Cortese}, L., et al. 2013, \mnras, 436,
	34
	
	\bibitem[{Chung} et al.(2009)]{Chung09}
	{Chung}, A., {van Gorkom}, J.~H., {Kenney}, J.~D.~P., {Crowl}, H., \&
	{Vollmer}, B. 2009, \aj, 138, 1741
	
	\bibitem[{Conroy} et al.(2009)]{Conroy09}
	{Conroy}, C., {Gunn}, J.~E., \& {White}, M. 2009, \apj, 699, 486
	
	\bibitem[{Conroy} et al.(2006)]{Conroy06}
	{Conroy}, C., {Wechsler}, R.~H., \& {Kravtsov}, A.~V. 2006, \apj, 647, 201
	
	\bibitem[{Crain} et al.(2017)]{Crain17}
	{Crain}, R.~A., {Bah{\'e}}, Y.~M., {Lagos}, C.~d.~P., et al. 2017, \mnras, 464,
	4204
	
	\bibitem[{Cunnama} et al.(2014)]{Cunnama14}
	{Cunnama}, D., {Andrianomena}, S., {Cress}, C.~M., et al. 2014, \mnras, 438,
	2530
	
	\bibitem[{Dalal} et al.(2008)]{Dalal08}
	{Dalal}, N., {White}, M., {Bond}, J.~R., \& {Shirokov}, A. 2008, \apj, 687, 12
	
	\bibitem[{Dav{\'e}} et al.(2013)]{Dave13}
	{Dav{\'e}}, R., {Katz}, N., {Oppenheimer}, B.~D., {Kollmeier}, J.~A., \&
	{Weinberg}, D.~H. 2013, \mnras, 434, 2645
	
	\bibitem[{Davis} \& {Peebles}(1983)]{Davis83}
	{Davis}, M., \& {Peebles}, P.~J.~E. 1983, \apj, 267, 465
	
	\bibitem[{Duffy} et al.(2012a)]{Duffy12}
	{Duffy}, A.~R., {Kay}, S.~T., {Battye}, R.~A., et al. 2012a, \mnras, 420, 2799
	
	\bibitem[{Duffy} et al.(2012b)]{Duffy12b}
	{Duffy}, A.~R., {Meyer}, M.~J., {Staveley-Smith}, L., et al. 2012b, \mnras,
	426, 3385
	
	\bibitem[{Evoli} et al.(2011)]{Evoli11}
	{Evoli}, C., {Salucci}, P., {Lapi}, A., \& {Danese}, L. 2011, \apj, 743, 45
	
	\bibitem[{Fern{\'a}ndez} et al.(2013)]{Fernandez13}
	{Fern{\'a}ndez}, X., {van Gorkom}, J.~H., {Hess}, K.~M., et al. 2013, \apjl,
	770, L29
	
	\bibitem[{Fern{\'a}ndez} et al.(2016)]{Fernandez16}
	{Fern{\'a}ndez}, X., {Gim}, H.~B., {van Gorkom}, J.~H., et al. 2016, \apjl,
	824, L1
	
	\bibitem[{Fixsen} et al.(1996)]{Fixsen96}
	{Fixsen}, D.~J., {Cheng}, E.~S., {Gales}, J.~M., et al. 1996, \apj, 473, 576
	
	\bibitem[{Fu} et al.(2010)]{Fu10}
	{Fu}, J., {Guo}, Q., {Kauffmann}, G., \& {Krumholz}, M.~R. 2010, \mnras, 409,
	515
	
	\bibitem[{Fu} et al.(2012)]{Fu12}
	{Fu}, J., {Kauffmann}, G., {Li}, C., \& {Guo}, Q. 2012, \mnras, 424, 2701
	
	\bibitem[{Fu} et al.(2013)]{Fu13}
	{Fu}, J., {Kauffmann}, G., {Huang}, M.-l., et al. 2013, \mnras, 434, 1531
	
	\bibitem[{Gao} et al.(2005)]{Gao05}
	{Gao}, L., {Springel}, V., \& {White}, S.~D.~M. 2005, \mnras, 363, L66
	
	\bibitem[{Gao} \& {White}(2007)]{Gao07}
	{Gao}, L., \& {White}, S.~D.~M. 2007, \mnras, 377, L5
	
	\bibitem[{Gavazzi} et al.(2005)]{Gavazzi05}
	{Gavazzi}, G., {Boselli}, A., {van Driel}, W., \& {O'Neil}, K. 2005, \aap, 429,
	439
	
	\bibitem[{Geach} et al.(2012)]{Geach12}
	{Geach}, J.~E., {Sobral}, D., {Hickox}, R.~C., et al. 2012, \mnras, 426, 679
	
	\bibitem[{Giovanelli} \& {Haynes}(1985)]{Giovanelli85}
	{Giovanelli}, R., \& {Haynes}, M.~P. 1985, \apj, 292, 404
	
	\bibitem[{Giovanelli} et al.(2005)]{Giovanelli05}
	{Giovanelli}, R., {Haynes}, M.~P., {Kent}, B.~R., et al. 2005, \aj, 130, 2598
	
	\bibitem[{Guo} et al.(2014a)]{Guo14}
	{Guo}, H., {Li}, C., {Jing}, Y.~P., \& {B{\"o}rner}, G. 2014a, \apj, 780, 139
	
	\bibitem[{Guo} et al.(2012)]{Guo12}
	{Guo}, H., {Zehavi}, I., \& {Zheng}, Z. 2012, \apj, 756, 127
	
	\bibitem[{Guo} et al.(2013)]{Guo13}
	{Guo}, H., {Zehavi}, I., {Zheng}, Z., et al. 2013, \apj, 767, 122
	
	\bibitem[{Guo} et al.(2014b)]{Guo14b}
	{Guo}, H., {Zheng}, Z., {Zehavi}, I., et al. 2014b, \mnras, 441, 2398
	
	\bibitem[{Guo} et al.(2015a)]{Guo15b}
	{Guo}, H., {Zheng}, Z., {Zehavi}, I., et al. 2015a, \mnras, 453, 4368
	
	\bibitem[{Guo} et al.(2015b)]{Guo15c}
	{Guo}, H., {Zheng}, Z., {Zehavi}, I., et al. 2015b, \mnras, 446, 578
	
	\bibitem[{Guo} et al.(2016)]{Guo16}
	{Guo}, H., {Zheng}, Z., {Behroozi}, P.~S., et al. 2016, \mnras, 459, 3040
	
	\bibitem[{Guo} et al.(2010)]{Guo10}
	{Guo}, Q., {White}, S., {Li}, C., \& {Boylan-Kolchin}, M. 2010, \mnras, 404,
	1111
	
	\bibitem[{Hamilton}(1992)]{Hamilton92}
	{Hamilton}, A.~J.~S. 1992, \apjl, 385, L5
	
	\bibitem[{Haynes} et al.(2011)]{Haynes11}
	{Haynes}, M.~P., {Giovanelli}, R., {Martin}, A.~M., et al. 2011, \aj, 142, 170
	
	\bibitem[{Hearin} \& {Watson}(2013)]{Hearin-Watson-13}
	{Hearin}, A.~P., \& {Watson}, D.~F. 2013, \mnras, 435, 1313
	
	\bibitem[{Hearin} et al.(2016)]{Hearin16}
	{Hearin}, A.~P., {Zentner}, A.~R., {van den Bosch}, F.~C., {Campbell}, D., \&
	{Tollerud}, E. 2016, \mnras, 460, 2552
	
	\bibitem[{Holwerda} et al.(2012)]{Holwerda12}
	{Holwerda}, B.~W., {Blyth}, S.-L., \& {Baker}, A.~J. 2012, in IAU Symposium,
	Vol. 284, The Spectral Energy Distribution of Galaxies - SED 2011, ed. R.~J.
	{Tuffs} \& C.~C. {Popescu}, 496--499
	
	\bibitem[{Huang} et al.(2012)]{Huang12}
	{Huang}, S., {Haynes}, M.~P., {Giovanelli}, R., \& {Brinchmann}, J. 2012, \apj,
	756, 113
	
	\bibitem[{Jing}(1998)]{Jing98}
	{Jing}, Y.~P. 1998, \apjl, 503, L9
	
	\bibitem[{Jing} et al.(2007)]{Jing07}
	{Jing}, Y.~P., {Suto}, Y., \& {Mo}, H.~J. 2007, \apj, 657, 664
	
	\bibitem[{Jones} et al.(2016)]{Jones16}
	{Jones}, M.~G., {Papastergis}, E., {Haynes}, M.~P., \& {Giovanelli}, R. 2016,
	\mnras, 457, 4393
	
	\bibitem[{Kaiser}(1987)]{Kaiser87}
	{Kaiser}, N. 1987, \mnras, 227, 1
	
	\bibitem[{Kim} et al.(2017)]{Kim17}
	{Kim}, H.-S., {Wyithe}, J.~S.~B., {Baugh}, C.~M., et al. 2017, \mnras, 465, 111
	
	\bibitem[{Klypin} et al.(2016)]{Klypin16}
	{Klypin}, A., {Yepes}, G., {Gottl{\"o}ber}, S., {Prada}, F., \& {He{\ss}}, S.
	2016, \mnras, 457, 4340
	
	\bibitem[{Knebe} et al.(2013)]{Knebe13}
	{Knebe}, A., {Pearce}, F.~R., {Lux}, H., et al. 2013, \mnras, 435, 1618
	
	\bibitem[{Koribalski}(2012)]{Koribalski12}
	{Koribalski}, B.~S. 2012, \pasa, 29, 359
	
	\bibitem[{Kravtsov} et al.(2004)]{Kravtsov04}
	{Kravtsov}, A.~V., {Berlind}, A.~A., {Wechsler}, R.~H., et al. 2004, \apj, 609,
	35
	
	\bibitem[{Lacerna} \& {Padilla}(2011)]{Lacerna11}
	{Lacerna}, I., \& {Padilla}, N. 2011, \mnras, 412, 1283
	
	\bibitem[{Lagos} et al.(2011)]{Lagos11b}
	{Lagos}, C.~D.~P., {Lacey}, C.~G., {Baugh}, C.~M., {Bower}, R.~G., \& {Benson},
	A.~J. 2011, \mnras, 416, 1566
	
	\bibitem[{Landy} \& {Szalay}(1993)]{Landy93}
	{Landy}, S.~D., \& {Szalay}, A.~S. 1993, \apj, 412, 64
	
	\bibitem[{Leauthaud} et al.(2012)]{Leauthaud12}
	{Leauthaud}, A., {Tinker}, J., {Bundy}, K., et al. 2012, \apj, 744, 159
	
	\bibitem[{Li} et al.(2009)]{Li09}
	{Li}, C., {Gadotti}, D.~A., {Mao}, S., \& {Kauffmann}, G. 2009, \mnras, 397,
	726
	
	\bibitem[{Li} et al.(2006)]{Li06}
	{Li}, C., {Jing}, Y.~P., {Kauffmann}, G., et al. 2006, \mnras, 368, 37
	
	\bibitem[{Li} et al.(2012)]{Li12}
	{Li}, C., {Kauffmann}, G., {Fu}, J., et al. 2012, \mnras, 424, 1471
	
	\bibitem[{Li} et al.(2008)]{Li08}
	{Li}, Y., {Mo}, H.~J., \& {Gao}, L. 2008, \mnras, 389, 1419
	
	\bibitem[{Lin} et al.(2016)]{Lin16}
	{Lin}, Y.-T., {Mandelbaum}, R., {Huang}, Y.-H., et al. 2016, \apj, 819, 119
	
	\bibitem[{Lu} et al.(2014)]{LuZ14}
	{Lu}, Z., {Mo}, H.~J., {Lu}, Y., et al. 2014, \mnras, 439, 1294
	
	\bibitem[{Lu} et al.(2015)]{LuZ15}
	{Lu}, Z., {Mo}, H.~J., {Lu}, Y., et al. 2015, \mnras, 450, 1604
	
	\bibitem[{Luo} et al.(2016)]{Luo16}
	{Luo}, Y., {Kang}, X., {Kauffmann}, G., \& {Fu}, J. 2016, \mnras, 458, 366
	
	\bibitem[{Magri} et al.(1988)]{Magri88}
	{Magri}, C., {Haynes}, M.~P., {Forman}, W., {Jones}, C., \& {Giovanelli}, R.
	1988, \apj, 333, 136
	
	\bibitem[{Mandelbaum} et al.(2016)]{Mandelbaum16}
	{Mandelbaum}, R., {Wang}, W., {Zu}, Y., et al. 2016, \mnras, 457, 3200
	
	\bibitem[{Martin} et al.(2012)]{Martin12}
	{Martin}, A.~M., {Giovanelli}, R., {Haynes}, M.~P., \& {Guzzo}, L. 2012, \apj,
	750, 38
	
	\bibitem[{Martin} et al.(2010)]{Martin10}
	{Martin}, A.~M., {Papastergis}, E., {Giovanelli}, R., et al. 2010, \apj, 723,
	1359
	
	\bibitem[{McGaugh}(2012)]{McGaugh12}
	{McGaugh}, S.~S. 2012, \aj, 143, 40
	
	\bibitem[{Meyer}(2009)]{Meyer09}
	{Meyer}, M. 2009, in Panoramic Radio Astronomy: Wide-field 1-2 GHz Research on
	Galaxy Evolution, 15
	
	\bibitem[{Meyer} et al.(2007)]{Meyer07}
	{Meyer}, M.~J., {Zwaan}, M.~A., {Webster}, R.~L., {Brown}, M.~J.~I., \&
	{Staveley-Smith}, L. 2007, \apj, 654, 702
	
	\bibitem[{Meyer} et al.(2004)]{Meyer04}
	{Meyer}, M.~J., {Zwaan}, M.~A., {Webster}, R.~L., et al. 2004, \mnras, 350,
	1195
	
	\bibitem[{Miyatake} et al.(2016)]{Miyatake16}
	{Miyatake}, H., {More}, S., {Takada}, M., et al. 2016, Physical Review Letters,
	116, 041301
	
	\bibitem[{More} et al.(2016)]{More16}
	{More}, S., {Miyatake}, H., {Takada}, M., et al. 2016, \apj, 825, 39
	
	\bibitem[{Moster} et al.(2010)]{Moster10}
	{Moster}, B.~P., {Somerville}, R.~S., {Maulbetsch}, C., et al. 2010, \apj, 710,
	903
	
	\bibitem[{Norberg} et al.(2011)]{Norberg11}
	{Norberg}, P., {Gazta{\~n}aga}, E., {Baugh}, C.~M., \& {Croton}, D.~J. 2011,
	\mnras, 418, 2435
	
	\bibitem[{Nuza} et al.(2013)]{Nuza13}
	{Nuza}, S.~E., {S{\'a}nchez}, A.~G., {Prada}, F., et al. 2013, \mnras, 432, 743
	
	\bibitem[{Obreschkow} et al.(2009)]{Obreschkow09}
	{Obreschkow}, D., {Croton}, D., {De Lucia}, G., {Khochfar}, S., \& {Rawlings},
	S. 2009, \apj, 698, 1467
	
	\bibitem[{Onions} et al.(2012)]{Onions12}
	{Onions}, J., {Knebe}, A., {Pearce}, F.~R., et al. 2012, \mnras, 423, 1200
	
	\bibitem[{Padmanabhan} \& {Kulkarni}(2017)]{Padmanabhan17}
	{Padmanabhan}, H., \& {Kulkarni}, G. 2017, \mnras, 470, 340
	
	\bibitem[{Padmanabhan} \& {Refregier}(2017)]{Padmanabhan17a}
	{Padmanabhan}, H., \& {Refregier}, A. 2017, \mnras, 464, 4008
	
	\bibitem[{Padmanabhan} et al.(2017)]{Padmanabhan17b}
	{Padmanabhan}, H., {Refregier}, A., \& {Amara}, A. 2017, \mnras, 469, 2323
	
	\bibitem[{Papastergis} et al.(2013)]{Papastergis13}
	{Papastergis}, E., {Giovanelli}, R., {Haynes}, M.~P.,
	{Rodr{\'{\i}}guez-Puebla}, A., \& {Jones}, M.~G. 2013, \apj, 776, 43
	
	\bibitem[{Papastergis} et al.(2011)]{Papastergis11}
	{Papastergis}, E., {Martin}, A.~M., {Giovanelli}, R., \& {Haynes}, M.~P. 2011,
	\apj, 739, 38
	
	\bibitem[{Peacock} \& {Smith}(2000)]{Peacock00}
	{Peacock}, J.~A., \& {Smith}, R.~E. 2000, \mnras, 318, 1144
	
	\bibitem[{Peebles}(1969)]{Peebles69}
	{Peebles}, P.~J.~E. 1969, \apj, 155, 393
	
	\bibitem[{Planck Collaboration}(2014)]{PlanckCollaboration14}
	{Planck Collaboration}. 2014, \aap, 571, A16
	
	\bibitem[{Popping} et al.(2009)]{Popping09}
	{Popping}, A., {Dav{\'e}}, R., {Braun}, R., \& {Oppenheimer}, B.~D. 2009, \aap,
	504, 15
	
	\bibitem[{Popping} et al.(2015)]{Popping15}
	{Popping}, G., {Behroozi}, P.~S., \& {Peeples}, M.~S. 2015, \mnras, 449, 477
	
	\bibitem[{Popping} et al.(2014)]{Popping14b}
	{Popping}, G., {Somerville}, R.~S., \& {Trager}, S.~C. 2014, \mnras, 442, 2398
	
	\bibitem[{Rafieferantsoa} et al.(2015)]{Rafieferantsoa15}
	{Rafieferantsoa}, M., {Dav{\'e}}, R., {Angl{\'e}s-Alc{\'a}zar}, D., et al.
	2015, \mnras, 453, 3980
	
	\bibitem[{Reddick} et al.(2013)]{Reddick13}
	{Reddick}, R.~M., {Wechsler}, R.~H., {Tinker}, J.~L., \& {Behroozi}, P.~S.
	2013, \apj, 771, 30
	
	\bibitem[{Rodr{\'{\i}}guez-Puebla} et al.(2013)]{Rodriguez-Puebla13}
	{Rodr{\'{\i}}guez-Puebla}, A., {Avila-Reese}, V., \& {Drory}, N. 2013, \apj,
	767, 92
	
	\bibitem[{Ross} et al.(2012)]{Ross12}
	{Ross}, A.~J., {Percival}, W.~J., {S{\'a}nchez}, A.~G., et al. 2012, \mnras,
	424, 564
	
	\bibitem[{Saintonge} et al.(2011)]{Saintonge11}
	{Saintonge}, A., {Kauffmann}, G., {Kramer}, C., et al. 2011, \mnras, 415, 32
	
	\bibitem[{Sawala} et al.(2015)]{Sawala15}
	{Sawala}, T., {Frenk}, C.~S., {Fattahi}, A., et al. 2015, \mnras, 448, 2941
	
	\bibitem[{Scoccimarro} et al.(2001)]{Scoccimarro01}
	{Scoccimarro}, R., {Sheth}, R.~K., {Hui}, L., \& {Jain}, B. 2001, \apj, 546, 20
	
	\bibitem[{Seljak}(2000)]{Seljak00}
	{Seljak}, U. 2000, \mnras, 318, 203
	
	\bibitem[{Shankar} et al.(2006)]{Shankar-06}
	{Shankar}, F., {Lapi}, A., {Salucci}, P., {De Zotti}, G., \& {Danese}, L. 2006,
	\apj, 643, 14
	
	\bibitem[{Skibba} et al.(2014)]{Skibba14}
	{Skibba}, R.~A., {Smith}, M.~S.~M., {Coil}, A.~L., et al. 2014, \apj, 784, 128
	
	\bibitem[{Skibba} et al.(2015)]{Skibba15}
	{Skibba}, R.~A., {Coil}, A.~L., {Mendez}, A.~J., et al. 2015, \apj, 807, 152
	
	\bibitem[{Solanes} et al.(2001)]{Solanes01}
	{Solanes}, J.~M., {Manrique}, A., {Garc{\'{\i}}a-G{\'o}mez}, C., et al. 2001,
	\apj, 548, 97
	
	\bibitem[{Springel} et al.(2005)]{Springel05b}
	{Springel}, V., {White}, S.~D.~M., {Jenkins}, A., et al. 2005, \nat, 435, 629
	
	\bibitem[{Steidel} et al.(2010)]{Steidel10}
	{Steidel}, C.~C., {Erb}, D.~K., {Shapley}, A.~E., et al. 2010, \apj, 717, 289
	
	\bibitem[{Teimoorinia} et al.(2017)]{Teimoorinig17}
	{Teimoorinia}, H., {Ellison}, S.~L., \& {Patton}, D.~R. 2017, \mnras, 464, 3796
	
	\bibitem[{Tinker} et al.(2005)]{Tinker05}
	{Tinker}, J.~L., {Weinberg}, D.~H., {Zheng}, Z., \& {Zehavi}, I. 2005, \apj,
	631, 41
	
	\bibitem[{Vale} \& {Ostriker}(2004)]{Vale-Ostriker-04}
	{Vale}, A., \& {Ostriker}, J.~P. 2004, \mnras, 353, 189
	
	\bibitem[{Vale} \& {Ostriker}(2006)]{Vale06}
	{Vale}, A., \& {Ostriker}, J.~P. 2006, \mnras, 371, 1173
	
	\bibitem[{van den Bosch} et al.(2013)]{vandenBosch13}
	{van den Bosch}, F.~C., {More}, S., {Cacciato}, M., {Mo}, H., \& {Yang}, X.
	2013, \mnras, 430, 725
	
	\bibitem[{Vogelsberger} et al.(2014)]{Vogelsberger14b}
	{Vogelsberger}, M., {Genel}, S., {Springel}, V., et al. 2014, \mnras, 444, 1518
	
	\bibitem[{Wang} et al.(2007a)]{Wang07}
	{Wang}, H.~Y., {Mo}, H.~J., \& {Jing}, Y.~P. 2007a, \mnras, 375, 633
	
	\bibitem[{Wang} et al.(2007b)]{WangL07}
	{Wang}, L., {Li}, C., {Kauffmann}, G., \& {De Lucia}, G. 2007b, \mnras, 377,
	1419
	
	\bibitem[{Wechsler} et al.(2006)]{Wechsler06}
	{Wechsler}, R.~H., {Zentner}, A.~R., {Bullock}, J.~S., {Kravtsov}, A.~V., \&
	{Allgood}, B. 2006, \apj, 652, 71
	
	\bibitem[{White} et al.(2011)]{White11}
	{White}, M., {Blanton}, M., {Bolton}, A., et al. 2011, \apj, 728, 126
	
	\bibitem[{Xie} et al.(2017)]{Xie17}
	{Xie}, L., {De Lucia}, G., {Hirschmann}, M., {Fontanot}, F., \& {Zoldan}, A.
	2017, \mnras, 469, 968
	
	\bibitem[{Xu} et al.(2016)]{Xu16}
	{Xu}, H., {Zheng}, Z., {Guo}, H., {Zhu}, J., \& {Zehavi}, I. 2016, \mnras, 460,
	3647
	
	\bibitem[{Yang} et al.(2004)]{Yang04}
	{Yang}, X., {Mo}, H.~J., {Jing}, Y.~P., {van den Bosch}, F.~C., \& {Chu}, Y.
	2004, \mnras, 350, 1153
	
	\bibitem[{Yang} et al.(2003)]{Yang03}
	{Yang}, X., {Mo}, H.~J., \& {van den Bosch}, F.~C. 2003, \mnras, 339, 1057
	
	\bibitem[{Yang} et al.(2012)]{Yang12}
	{Yang}, X., {Mo}, H.~J., {van den Bosch}, F.~C., {Zhang}, Y., \& {Han}, J.
	2012, \apj, 752, 41
	
	\bibitem[{York} et al.(2000)]{York00}
	{York}, D.~G., {Adelman}, J., {Anderson}, Jr., J.~E., et al. 2000, \aj, 120,
	1579
	
	\bibitem[{Zavala} et al.(2009)]{Zavala09}
	{Zavala}, J., {Jing}, Y.~P., {Faltenbacher}, A., et al. 2009, \apj, 700, 1779
	
	\bibitem[{Zehavi} et al.(2002)]{Zehavi02}
	{Zehavi}, I., {Blanton}, M.~R., {Frieman}, J.~A., et al. 2002, \apj, 571, 172
	
	\bibitem[{Zehavi} et al.(2005)]{Zehavi05}
	{Zehavi}, I., {Eisenstein}, D.~J., {Nichol}, R.~C., et al. 2005, \apj, 621, 22
	
	\bibitem[{Zehavi} et al.(2011)]{Zehavi11}
	{Zehavi}, I., {Zheng}, Z., {Weinberg}, D.~H., et al. 2011, \apj, 736, 59
	
	\bibitem[{Zentner} et al.(2016)]{Zentner16}
	{Zentner}, A.~R., {Hearin}, A., {van den Bosch}, F.~C., {Lange}, J.~U., \&
	{Villarreal}, A. 2016, ArXiv e-prints
	
	\bibitem[{Zentner} et al.(2014)]{Zentner14}
	{Zentner}, A.~R., {Hearin}, A.~P., \& {van den Bosch}, F.~C. 2014, \mnras, 443,
	3044
	
	\bibitem[{Zhang} et al.(2012)]{Zhang12}
	{Zhang}, H.-X., {Hunter}, D.~A., {Elmegreen}, B.~G., {Gao}, Y., \& {Schruba},
	A. 2012, \aj, 143, 47
	
	\bibitem[{Zhang} et al.(2013)]{Zhang13}
	{Zhang}, W., {Li}, C., {Kauffmann}, G., \& {Xiao}, T. 2013, \mnras, 429, 2191
	
	\bibitem[{Zhang} et al.(2009)]{Zhang09}
	{Zhang}, W., {Li}, C., {Kauffmann}, G., et al. 2009, \mnras, 397, 1243
	
	\bibitem[{Zheng} et al.(2007)]{Zheng07}
	{Zheng}, Z., {Coil}, A.~L., \& {Zehavi}, I. 2007, \apj, 667, 760
	
	\bibitem[{Zheng} et al.(2009)]{Zheng09}
	{Zheng}, Z., {Zehavi}, I., {Eisenstein}, D.~J., {Weinberg}, D.~H., \& {Jing},
	Y.~P. 2009, \apj, 707, 554
	
	\bibitem[{Zheng} et al.(2005)]{Zheng05}
	{Zheng}, Z., {Berlind}, A.~A., {Weinberg}, D.~H., et al. 2005, \apj, 633, 791
	
	\bibitem[{Zhu} et al.(2006)]{Zhu06}
	{Zhu}, G., {Zheng}, Z., {Lin}, W.~P., et al. 2006, \apjl, 639, L5
	
	\bibitem[{Zoldan} et al.(2017)]{Zoldan17}
	{Zoldan}, A., {De Lucia}, G., {Xie}, L., {Fontanot}, F., \& {Hirschmann}, M.
	2017, \mnras, 465, 2236
	
	\bibitem[{Zu} \& {Mandelbaum}(2015)]{Zu15}
	{Zu}, Y., \& {Mandelbaum}, R. 2015, \mnras, 454, 1161
	
	\bibitem[{Zu} \& {Mandelbaum}(2016)]{Zu16}
	{Zu}, Y., \& {Mandelbaum}, R. 2016, \mnras, 457, 4360
	
	\bibitem[{Zu} et al.(2016)]{Zu16b}
	{Zu}, Y., {Mandelbaum}, R., {Simet}, M., {Rozo}, E., \& {Rykoff}, E.~S. 2016,
	ArXiv e-prints
	
	\bibitem[{Zwaan} et al.(2005)]{Zwaan05}
	{Zwaan}, M.~A., {Meyer}, M.~J., {Staveley-Smith}, L., \& {Webster}, R.~L. 2005,
	\mnras, 359, L30
	
\end{thebibliography}
\end{document}